\documentclass[aps,prc,superscriptaddress,twoside,twocolumn,nofootinbib,showpacs]{revtex4}
\usepackage{amsmath,amssymb}
\usepackage{mathtools}

\usepackage{bbold}
\usepackage[usenames, dvipsnames]{xcolor}
\usepackage{braket}
\usepackage{graphicx,bm}
\usepackage{slashed}
\usepackage{booktabs}
\usepackage{tensor}

\usepackage{multirow}
\usepackage{mwe}

\usepackage{changes}

\usepackage{enumerate}
\definechangesauthor[name={Yeunhwan}, color=magenta]{YH}

\allowdisplaybreaks

\newcommand{\msun}{M_{\odot}}

\newcommand{\pt}{\partial}

\begin{document}

\title{Sensitivity of neutron drip lines and neutron star properties to the symmetry
energy}


\author{Yeunhwan \surname{Lim} }
\email{ylim@yonsei.ac.kr}
\affiliation{Department of Physics, Yonsei University, Seoul 03722, South Korea}

\author{Jeremy~W. Holt}
 \email{holt@physics.tamu.edu}
\affiliation{Cyclotron Institute, Texas A\&M University, College Station, TX 77843, USA}
\affiliation{Department of Physics and Astronomy, Texas A\&M University, College Station, TX 77843, USA }

\date{\today}

\begin{abstract}

We investigate the influence of the nuclear symmetry energy and its density slope parameter on the neutron dripline and neutron star properties using a semi-classical liquid drop model (LDM) and energy density functionals constrained by chiral effective field theory. To analyze finite nuclei and mass tables, the nuclear symmetry energy at saturation density is fixed, and the surface tension is determined to minimize the root-mean-square deviation of the total binding energy for 2208 nuclei. Correlations between symmetry energy parameters and neutron driplines, crust-core transition densities, and the radii of $1.4\,\msun$ neutron stars are explored using the LDM framework. Additionally, we examine the relationship between macroscopic properties, such as neutron star radii ($R_{1.4}$), and microscopic properties, including the number of isotopes and the last bound nucleus for $Z=28$, within the LDM context.
\end{abstract}


\maketitle


\section{Introduction}

The nuclear symmetry energy, defined as the difference between the energy per particle of homogeneous neutron matter and symmetric nuclear matter at a given density, is an important organizing concept in nuclear structure physics that helps to explain nuclear binding energies\,\cite{Myers69,Lattimer12a,Pearson2014}, neutron skin thicknesses\,\cite{chen05,centelles09,Roca2011}, and the location of the neutron dripline\,\cite{oyamatsu10,wang15} far from nuclear stability. The symmetry energy also plays a role in the dynamics of nuclei, including $r$-process nucleosynthesis~\cite{wang15}, the lifetime of heavy nuclei through $\alpha$ decay~\cite{dong11,shin16,lim17alfa}, and heavy-ion collisions~\cite{Li2002a,Li2002b, Tsang2004, Di2010}. It is crucial for understanding the properties of neutron stars as well. In particular, it governs the proton fraction as a function of the baryon density, neutron star cooling processes~\cite{page1992,Lim2017a}, the thickness of the crust, and the typical radii of neutron stars~\cite{lattimer01,gandolfi2012,Lattimer12a,steiner2012a,lim2019moi,lim19e}. For these reasons the nuclear symmetry energy is a prime focus of experimental investigations at rare-isotope beam facilities such as the Radioactive Isotope Beam Factory (RIBF) at RIKEN, the Facility for Antiproton and Ion Research (FAIR) at GSI, and the Facility for Rare Isotope Beams (FRIB) at MSU.

The density dependence of the nuclear symmetry energy is often encoded in a Taylor series expansion about nuclear matter saturation density ($n_0 \sim 0.16\,\mathrm{fm}^{-3}$), where $S_v$ is the symmetry energy at nuclear saturation density, $L$ is the slope parameter, $K_{\rm{sym}}$ is the curvature, and $Q_{\rm{sym}}$ is the skewness. Correlations among these empirical parameters can be studied through systematic investigations of finite nuclei properties, such neutron skin thicknesses and nuclear mass models\,\cite{baldo2016a,newton2021a,lattimer2023a}, microscopic calculations of the nuclear matter equation of state \cite{hebeler11,gezerlis13,roggero14,wlazlowski14,drischler14,drischler15,drischler16,tews16,Holt:2016pjb}, or the isovector quasiparticle interaction from Fermi liquid theory \cite{Holt:2011yj,holt2018b}. More recently, gravitational wave observations \cite{gw170817,gw170817apj,de18,ligo18a} have been utilized to study nuclear symmetry energy correlations\,\cite{fattoyev18,Krastev2018,Lim2018,Zhang2018a,Zhang2018b,lim2019moi,lim19e}. 

In the present work, we perform a joint investigation of finite nuclei across the nuclear chart and the properties of neutron stars using the nuclear symmetry energy as a common point of reference. Such an investigation can be conveniently carried out with semi-classical methods, such as the liquid drop model. In particular, our goal is to understand how the symmetry energy and neutron dripline are correlated with properties of neutron stars, such as the crust-core transition density or radius. In the compressible liquid drop model, we explore a range of scenarios by first fixing the nuclear symmetry energy and its slope parameter and then fitting the remaining free parameters to minimize the root-mean-square deviation with nuclear binding energies. 
Finally, we investigate the properties of $1.4\,M_\odot$ neutron stars in terms of the nuclear symmetry energy and correlations with the neutron dripline for specific isotopes.

The paper is organized as follows. In Section~\ref{sec:ldm} we use the LDM, including nuclear shell corrections, to calculate the total binding energy of finite nuclei and the neutron dripline. 
In Section \ref{results} we demonstrate that global properties of heavy nuclei in neutron star crusts as well as the neutron star radius depend on the nuclear symmetry energy parameters and properties of the neutron dripline. We discuss and summarize our results in Section~\ref{sec:con}.

\section{Liquid drop model}
\label{sec:ldm}
The liquid drop model (LDM) originates from the semi-empirical mass formula, which was developed to describe the masses of finite nuclei. Despite its simplicity, the most accurate nuclear mass models to date - such as the Myers models, DZ, FRDM, and WS4 - are still based on the liquid drop framework \cite{Myers1966, DZ1995, frdm12, wang2014}.
In this section, we use simplified liquid drop models to explore the properties of finite nuclei and to investigate the correlation between the symmetry energy and the location of neutron driplines.

\subsection{Incompressible liquid drop model and finite nuclei}
The importance of the nuclear symmetry energy was first recognized in the context of the incompressible LDM. In the LDM, the total binding energy  for a given nucleus is given by~\cite{Myers69,Steiner2005325}:
\begin{equation}\label{eq:incompldm}
\begin{aligned}
E(Z,A) & = -BA + E_S A^{2/3} + \frac{S_v}{1 + \frac{S_s}{S_v}A^{-1/3}} A I^2 \\
& + E_C \frac{Z^2}{A^{1/3}}  + E_{\rm ex}\frac{Z^{4/3}}{A^{1/3}}
+ E_{\rm diff}\frac{Z^2}{A} \\
& + E_{\rm{pair}} + E_{\rm{shell}},
\end{aligned}
\end{equation}
where $B$ is the binding energy of bulk nuclear matter, $E_S$ is the surface energy term, $S_v$ is the bulk symmetry energy, $S_s$ is the surface symmetry energy, $E_C$ is the Coulomb energy term, $E_{\rm ex}$ is the exchange Coulomb energy, and $E_{\rm diff}$ is the Coulomb diffuseness energy from the surface of the nucleus\,\cite{Myers1966}. 
The diffuseness energy 
accounts for the fact that the proton distribution in the nucleus is not a sharp surface, but rather a diffuse distribution.

We include a simple pairing energy contribution given by $E=a_p \frac{\Delta}{\sqrt{A}}$, where $a_p$ is $-1$ for even-even nuclei, $0$ for even-odd nuclei, and $+1$ for odd-odd nuclei. We adopt the method for shell energy corrections proposed by Duflo and Zuker~\cite{DZ1995,DI2009}. Here the shell energy contribution is given by
\begin{equation}
\label{shellcor}
E_{\text{shell}} = a_1 S_2 + a_2(S_2)^2 + a_3 S_3 + a_{nz}S_{nz},
\end{equation}
where the $a$ coefficients are found by minimizing the root mean square deviation in the total binding energy, and the $S$ functions
\begin{equation}
\begin{aligned}
S_2 & = \frac{n_v \bar{n}_v}{D_n} + \frac{z_v \bar{z}_v}{D_z}, \\
S_3 & = \frac{n_v \bar{n}_v(n_v-\bar{n}_v)}{D_n} + \frac{z_v \bar{z}_v (z_v - \bar{z}_v)}{D_z}, \\
S_{nz} & = \frac{n_v \bar{n}_v z_v \bar{z}_v}{D_n D_z},
\label{shells}
\end{aligned}
\end{equation}
depend on the neutron (proton) valence number $n_v (z_v)$, its complement $\bar{n}_v (\bar{z}_v)$, and the magic number difference between shells $D_n (D_z)$.

In the incompressible liquid drop model, the Coulomb energy contribution is obtained by assuming a uniform density in the nucleus. The charge radius of a nucleus in the LDM is not clear, since the root mean square radius is not linear in $A^{1/3}$ because of the compression of the density and the presence of neutron or proton skins on the surface. We assume that the charge radius is identical with the matter radius, $r = r_0A^{1/3}$. From this, we obtain the expressions for 
$E_C$, $E_{\rm ex}$, and $E_{\rm diff}$:
\begin{equation}
\begin{aligned}
& E_C \frac{Z^2}{A^{1/3}} = \frac{3Z^2e^2}{5r}\,,
\,\,\,
E_{\rm diff}\frac{Z^2}{A} = -\frac{\pi^2Z^2e^2 d^2}{2r^3}\,,
\\
& E_{\rm ex}\frac{Z^{4/3}}{A^{1/3}} =
-\frac{3Z^{4/3} e^2}{4r} \left( \frac{3}{2\pi}\right)^{2/3}\,,
\end{aligned}
\end{equation}
where $d=0.55$~fm~\cite{Steiner2005325,antonov05, hatakeyama18, choudhary21} is the surface diffuseness parameter. If each of these Coulomb coefficients is treated as a free variable in fitting to the known binding energies  of nuclei, the consistency of the Coulomb interaction terms is lost and does not enable a proper constraint on the nuclear symmetry energy. We set the saturation density $n_0 =0.155$~fm$^{-3}$ for the LDM in this work, which then gives $r_0 = \left( \frac{3n_0}{4\pi} \right)^{1/3}=1.155$~fm.
\begin{table*}[htp]
	\caption{Liquid drop model parameters that minimize the root mean square deviation in the total binding energy.}
	\label{tb:ldm}
	\centering
	\begin{tabular}{ccccccc}
		\hline
		\hline
		\\[-2.8ex]
		$B$~(MeV)     & $E_S$~(MeV) & $S_v$~(MeV) & $S_s/S_v$ & $\Delta$~(MeV) & RMSD (MeV)  & - \\
		\hline   
		15.893        & 20.212      & 28.699     &    1.846  & 12.935         & 2.620       & w/o shell\\
		\hline
		15.879       & 19.732      & 31.152      &    2.389  & 12.111        & 0.949       & w/ shell\\
		\hline
	\end{tabular}
\end{table*}
The parameters of the LDM can be obtained from $\chi^2$ minimization
\begin{equation}
\chi^2 = \frac{1}{N} \sum_{(A,Z)} \bigl[ E(A,Z)^{\text{Exp.}} - E(A,Z)^{\text{LDM}} \bigr]^2\, ,
\end{equation}
where we use the binding energies $B$ of 2208 nuclei~\cite{ame2016} in which $Z \ge 20$ and $A \ge 36$.

Table \ref{tb:ldm} shows the LDM parameters that minimize the root mean square deviation (RMSD) of the total binding energy. We show the results both with and without the inclusion of shell corrections. Note that the inclusion of shell corrections reduces the RMSD by more than a factor of 2. Compared to the most accurate mass model calculations\,\cite{DZ1995,frdm12,hfb24,wang2014}, this simple LDM approach is an accurate and practical tool to calculate nuclear masses for the general case\,\cite{lim17alfa,DI2009}.


From the LDM, we find a strong correlation between $S_v$ and $S_s/S_v$ of the following form:
\begin{equation}
\begin{aligned}\label{eq:corsssv}
S_s/S_v = & -5.111 + 0.244S_v \,\mathrm{MeV}^{-1} \\
&  \quad \text{(not including shell corrections) and} \\
S_s/S_v = & -5.266 + 0.246S_v \,\mathrm{MeV}^{-1}  \\
& \quad \text{(including shell corrections)}.
\end{aligned}
\end{equation}
Our results for the slope of the $S_s/S_v$ vs.\ $S_v$ correlation are in agreement with a previous work\,\cite{Steiner2005325} that obtained a similar correlation and slope from the LDM modified through the inclusion of neutron and proton skins. A strong correlation between $S_v$ and $S_s$ is expected since they both stem from the neutron and proton asymmetry.

\begin{figure}[t]
\includegraphics[scale=0.45]{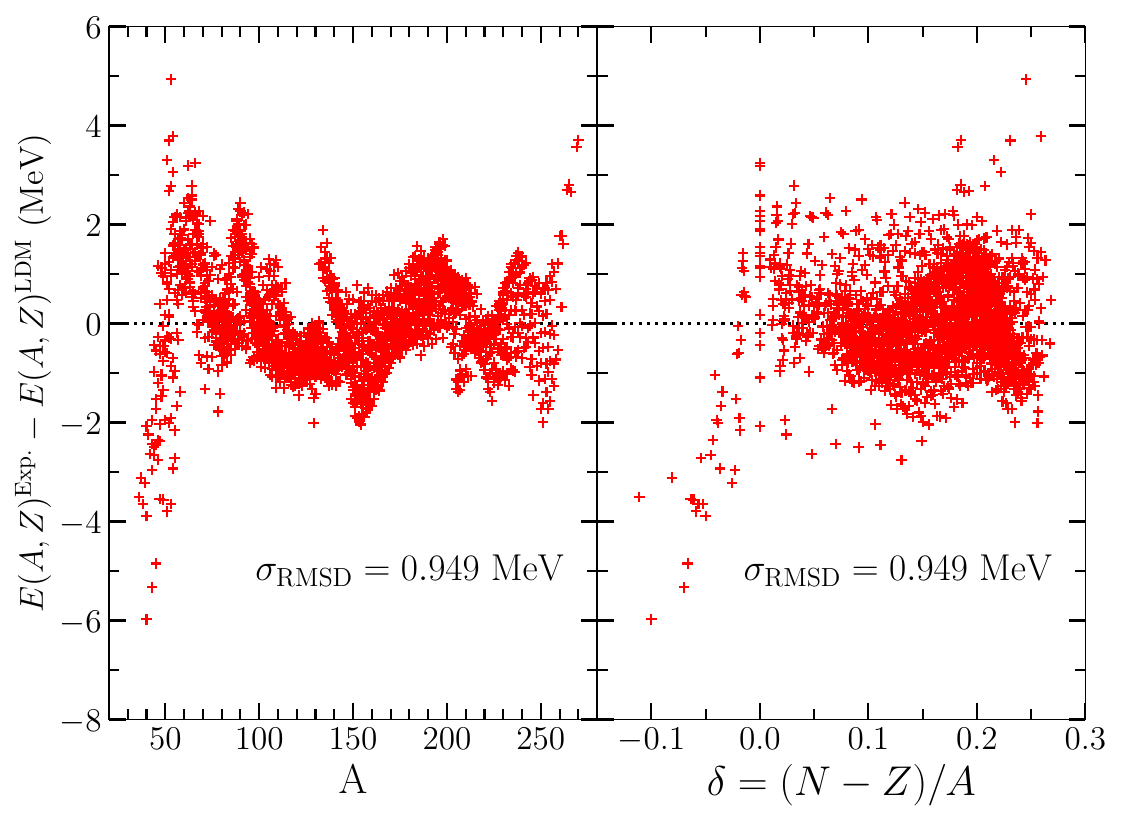}
\caption{Total binding energy difference between the LDM and experiment for given mass number (left) and isospin asymmetry (right).}
\label{fig:berror}
\end{figure}

Figure \ref{fig:berror} shows the difference between the LDM predictions for nuclear binding energies and experiment as a function of the mass number (left) and isospin asymmetry (right). Note that the LDM model does not provide an especially good description of proton-rich light nuclei, where the energy difference becomes large. This is partly caused by the fact that the radii of nuclei, given by $r =  r_0 A^{1/3}$ in the LDM, is estimated to be larger than that from Hartree-Fock (HF) calculations. In HF calculations, the central density of light nuclei is generally greater than nuclear matter saturation density. However, the central density in the incompressible LDM is constant for all nuclei. Thus, the Coulomb energy contribution to the total binding energy is expected to be lower than that from realistic density functional calculations. Note that the root mean square deviation is only 0.949~MeV in our global parameterization for 2208 nuclei. This indicates that for the study of proton-rich light nuclei, it is necessary to use density functional theory\,\cite{geng04,stoitsov05,goriely09} or an ab initio many-body techniques \cite{jdholt13,Hagen14,Hergert16}. Nevertheless, Figure \ref{fig:berror} demonstrates that the LDM gives quite
accurate results concerning the total binding energy for average (medium-mass and heavy) nuclei.

\begin{figure*}
	\begin{tabular}{c}
	\includegraphics[scale=0.253]{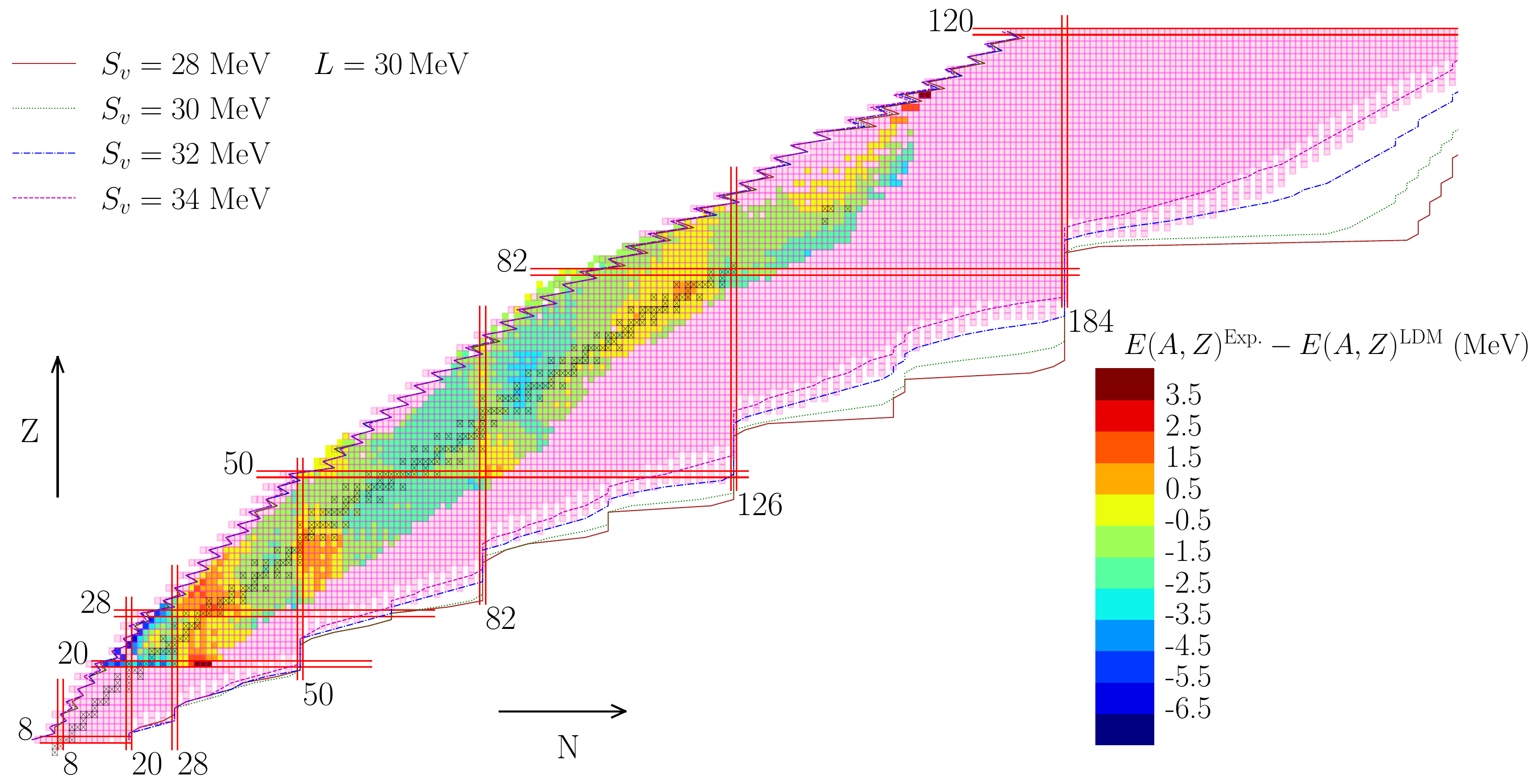} \\
	\includegraphics[scale=0.253]{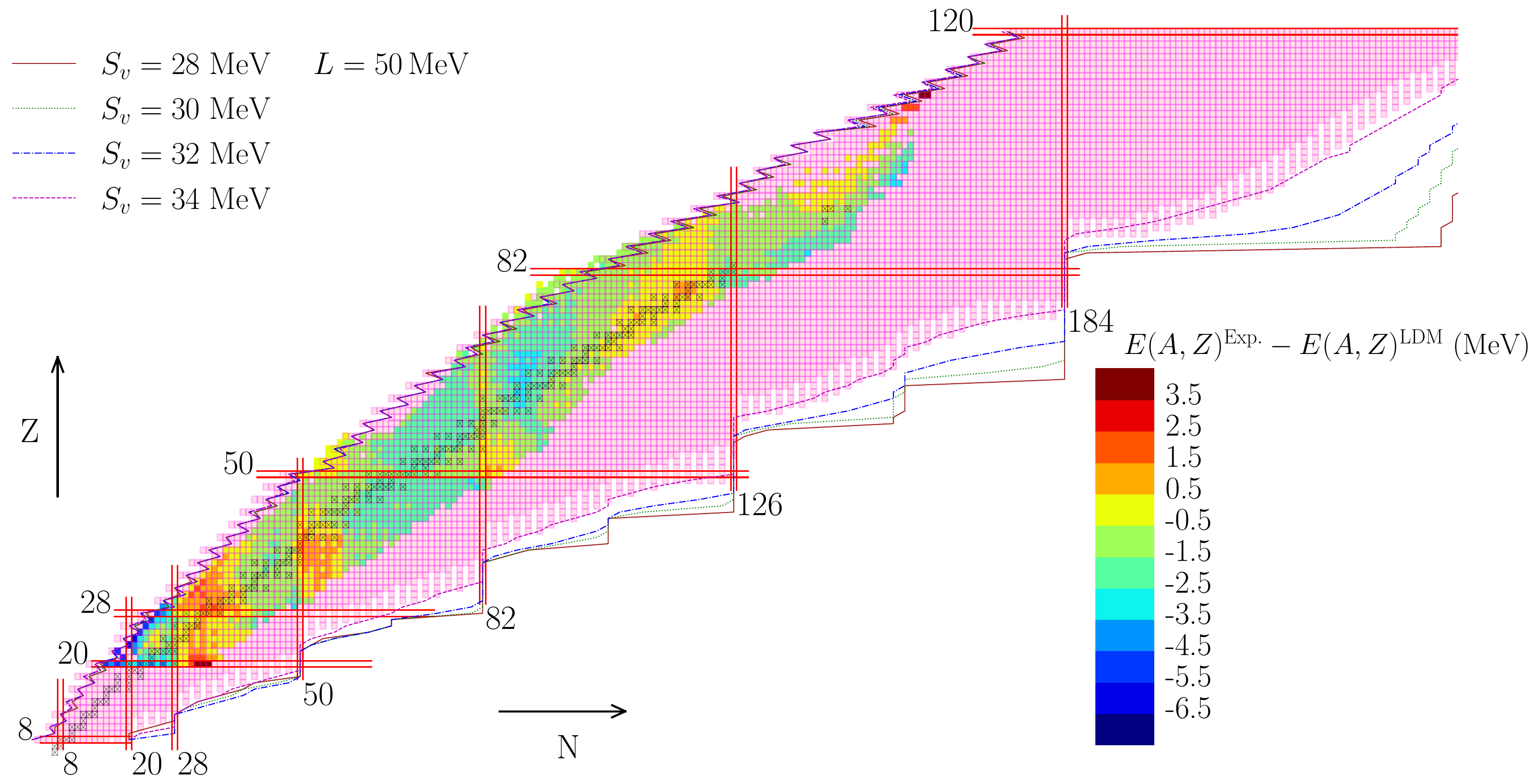} \\
	\includegraphics[scale=0.253]{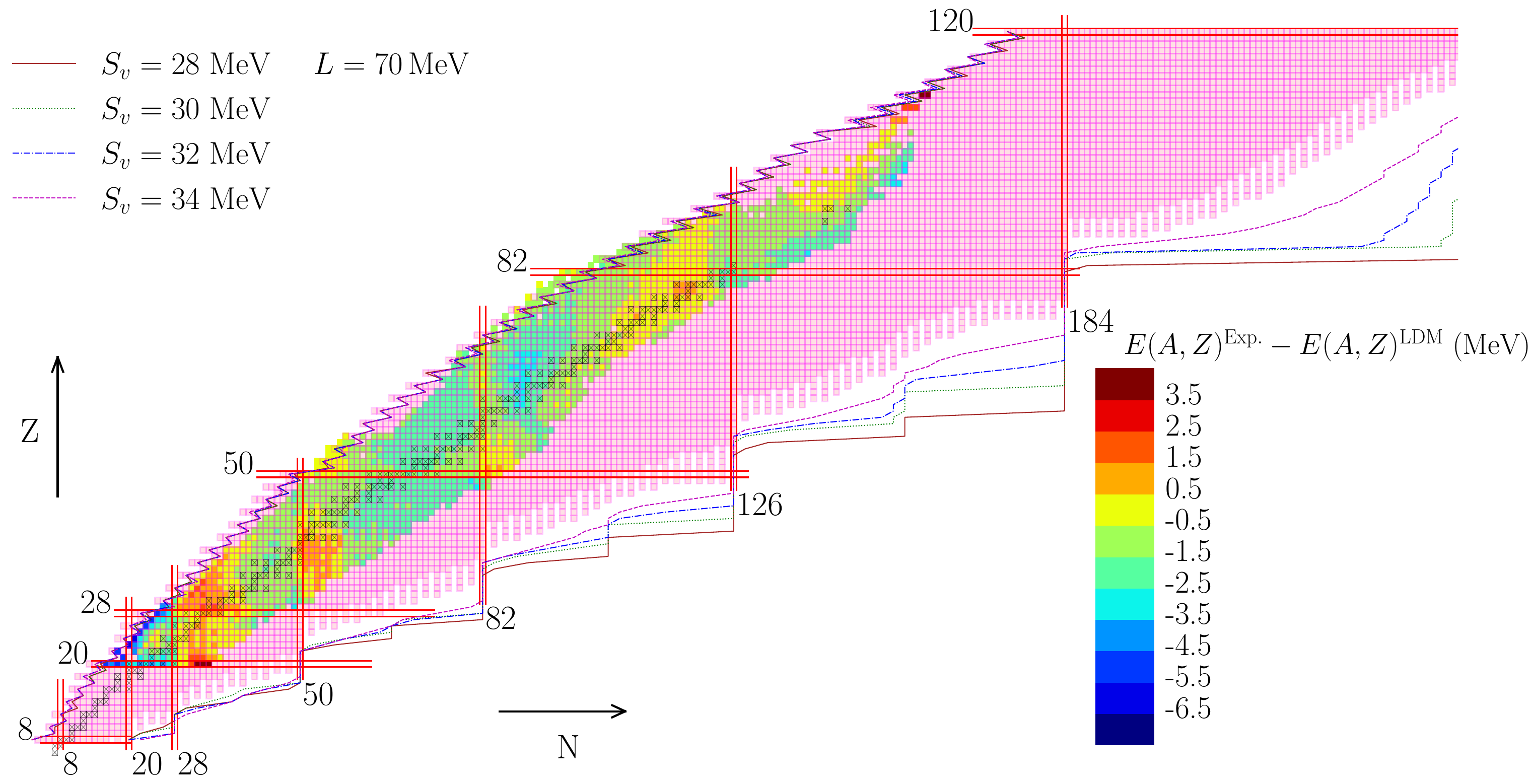} 
	\end{tabular}
	\caption{Neutron drip lines constructed from the incompressible LDM (pink squares) and the compressible liquid drop model (curves) up to $Z=120$ for fixed $L$ and varying $S_v$.}
	\label{fig:masstable}
\end{figure*}
\begin{figure*}
   \includegraphics[scale=0.253]{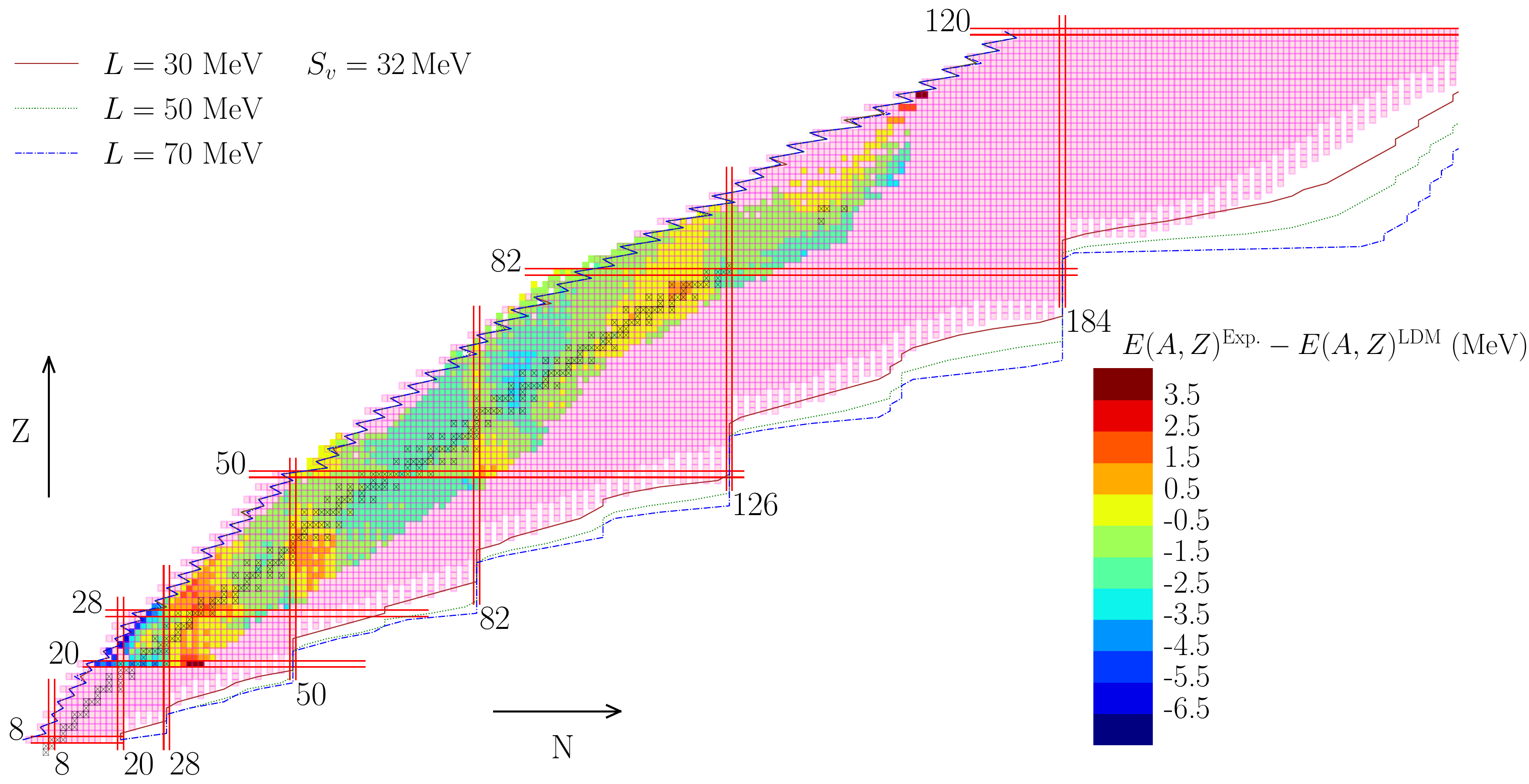} 
 \caption{Neutron drip lines constructed from the incompressible LDM (pink squares) and the compressible liquid drop model (curves) up to $Z=120$ for fixed $S_v$ and varying $L$.}
	\label{fig:mtbsv}
\end{figure*}
This elementary liquid drop model predicts that there are 8425 bound nuclei with $8 \le Z \le 119$. This estimate suggests that there will be around 5000 new nuclei that can be synthesized or explored in the laboratory. 
In this work, bound nuclei are identified by requiring that the four conditions $S_n > 0$, $S_p > 0$, $S_{2n} > 0$, and $S_{2p} > 0$ are satisfied across a sufficiently large set of $(A,Z)$ combinations. The neutron dripline is then defined as the last bound nucleus for each atomic number $Z$.
That is, 
the neutron dripline is determined by comparing the total binding energy between neighboring nuclei ($S_n < 0$), 
\begin{equation*}
E(Z,A+1) - E(Z,A) < 0\,,
\end{equation*}
where the binding energy is defined with a negative sign. 
Then, the energy difference caused by adding one more neutron in a nucleus can be approximated by
\begin{equation}\label{eq:ndrip}
\begin{aligned}
E(Z, & A+1)  - E(Z,A)  = \frac{\delta E}{\delta N}\cdot\{(N+1) - N\} + \frac{\Delta}{\sqrt{A}}\\
& \simeq - B + \frac{1}{3A} \left( 2E_S A^{2/3} - E_C\frac{Z^2}{A^{1/3}}\right) \\
& \quad + \frac{S_v}{1 + \frac{S_s}{S_v}A^{-1/3}}(2I - I^2) + \frac{\Delta }{\sqrt{A}}\,,
\end{aligned}
\end{equation}
where we neglect the exchange and diffuseness Coulomb energy since together they only account for $\sim 10\%$ of the classical Coulomb energy. We assume that the variation $S_v/(1+ S_sA^{-1/3}/S_v)$ is small in the above equation, since $1+a x^{-1/3} \simeq 1 + a (x+1)^{-1/3}$, where in this case $a \sim 1$ and $A > 40$.
Solving for $I$ in Eq.\ \eqref{eq:ndrip}, we then obtain the mass number $A$ that defines the neutron dripline for a given value of the atomic number $Z$:
\begin{equation}\label{eq:ndripsol}
A  \simeq 
\frac{2Z}
{ \left[ 
1  - \frac{1}{S_v}
\left(1 + \frac{S_s/S_v}{A^{1/3}}\right)
\left(B - \frac{\Delta}{\sqrt{A}} 
 + \frac{2E_S}{3A^{1/3}} - \frac{E_C Z^2}{3A^{4/3}}
\right)
\right]^{1/2}
}
\end{equation}
From the above equation, one observes the intuitive fact that as the symmetry energy increases, the dripline value of $A$ for a given atomic number $Z$ decreases. This is due to the fact that for higher symmetry energies, the binding energy decreases more rapidly as the neutron excess increases. We note that the above discussion is useful to obtain a closed expression for the dripline location as a function of the symmetry energy, but in our actual calculations of the dripline we do not employ approximations such as Eq.\ \eqref{eq:ndripsol}.

\subsection{Compressible liqud drop model and neutron driplines}
Compared to the incompressible liquid drop model (LDM), the compressible LDM accounts for variations in the central density of nuclei. As shown in Hartree-Fock and Hartree-Fock-Bogoliubov calculations \cite{Bennaceur05}, lighter nuclei tend to exhibit higher central densities than heavier ones. Moreover, the incompressible LDM does not accommodate changes in nuclear matter properties, as its parameters are fixed to minimize the root-mean-square deviation in the binding energies of finite nuclei. 
That is, the bulk matter properties--such as the binding energy per nucleon, symmetry energy, and saturation density--can be deduced only from the incompressible model parameters, as shown in Eq.~\eqref{eq:incompldm}, and there are no variations in these parameters.
Thus, to study the effects of both the nuclear symmetry energy $S_v$ and its slope parameter $L$, 
it is useful to employ the compressible LDM, which allows for the description of neutron skins that are necessary for a more accurate investigation of neutron-rich nuclei. 
In the compressible LDM, one takes the following functional form for the total binding energy\,\cite{Steiner2005325} :
\begin{equation}\label{eq:compldm}
\begin{aligned}
E(A,Z) & = f_B(n,x)(A -N_s) + 4\pi r^2 \sigma(\mu_n) + N_s \mu_n \\
& + E_{\rm Coul}
   + E_{\rm pair} + E_{\rm shell}, 
\end{aligned}
\end{equation}
where $N_s$ is the number of neutrons on the surface of the nucleus, $\mu_n$ is the chemical potential for neutrons on the surface of a nucleus,
$\sigma$ is a surface tension as a function of a neutron chemical potential, $\mu_n$, 
$f_B(n,x)$ is the energy per baryon for bulk nuclear matter
as a function of baryon number density $n$ and proton fraction $x$
\,\cite{Oyamatsu2007},
\begin{equation}\label{eq:fbldm}
\begin{aligned}
f_B(n,x)  & = -B + (1-2x)^2\left[ S_v + \frac{L}{3n_0}(n-n_0) \right] \\
& + \frac{K}{18n_0^2}(n-n_0)^2\,,
\end{aligned}
\end{equation}
and $E_{\rm Coul}$ is the Coulomb energy including exchange and diffuseness effects
in terms of radius $r$ and atomic number $Z$:
\begin{equation}
E_{\rm Coul} = 
\frac{3}{5}\frac{Z^2e^2}{r} 
- \frac{3Z^{4/3}e^2}{4r}\left(\frac{3}{2\pi} \right)^{2/3}
- \frac{\pi^2 Z^2 e^2d^2}{2r^3} .
\end{equation}
To minimize the total binding energy, we apply the Lagrange multiplier method with mass number and atomic number
\begin{equation}
A = \frac{4\pi}{3}r^3 n + N_s\,,\quad Z = (A-N_s)x\,
\end{equation}
as constraints. For given $A$ and $Z$, we have three unknowns $(n,x,r)$ and three equations,
\begin{subequations}
\begin{equation}
\begin{aligned}
 \frac{8\pi}{3}r \sigma(x)
+ \frac{\pt E_{\rm Coul}}{\pt r} -4\pi r^2 n^2 \frac{\pt f_B}{\pt n}= 0\,,
\end{aligned}
\end{equation}
\begin{equation}
A-N_s - \frac{4\pi}{3}r^3n  =0\,,
\end{equation}
\begin{equation}
Z - x(A-N_s) =0\,,
\end{equation}
\end{subequations}
where $\mu_n$ and $N_s$ are given as
\begin{equation}
\begin{aligned}
\mu_n & = f_B + n \frac{\pt f_B}{\pt n} - x \frac{\pt f_B}{\pt x}\,, \\
N_s    & = -4\pi r^2 \frac{\pt \sigma/\pt x}{\pt \mu_n /\pt x}\,.
\end{aligned}
\end{equation}
The surface tension as a function of proton fraction can be approximated by \cite{RPL1983,ls1991,lim2017c}:
\begin{equation}\label{eq:surfx}
\sigma(x) = \sigma_0 \frac{2\cdot 2^{\alpha} + q}{x^{-\alpha} + q +(1-x)^{-\alpha}}\,,
\end{equation}	
where $\sigma_0=\sigma(x=1/2)$. Then $\alpha$ and $q$ are fitting parameters for the surface tension.

\subsection{Compressible LDM parametrizations and results}

\begin{table}[h]
\caption{The RMSD for a given $(S_v, L)$ compressible liquid drop models. 
The number of isotopes for the Nickel ($Z=28$) added for comparison.}
    \centering
    \begin{tabular}{cccc}
    \hline
    \hline
       $S_v$\,(MeV) &  $L$ \,(MeV) & RMSD\,(MeV) & $\#$ of iso. of $Z=28$ \\
       \hline
         $28$     &  $30$     &  $1.505$     &    45    \\
         $30$     &  $30$     &  $1.264$     &    39  \\
         $32$     &  $30$     &  $1.110$     &    39  \\
         $34$     &  $30$     &  $1.066$     &    36   \\
      \hline      
         $28$     &  $50$     &  $1.617$     &   47   \\
         $30$     &  $50$     &  $1.321$     &   48   \\
         $32$     &  $50$     &  $1.104$     &   41   \\
         $34$     &  $50$     &  $1.004$     &   39   \\
      \hline      
         $28$     &  $70$     &  $1.784$     &   48    \\
         $30$     &  $70$     &  $1.446$     &   48   \\
         $32$     &  $70$     &  $1.180$     &   48   \\
         $34$     &  $70$     &  $1.013$     &   41   \\
      \hline          
    \end{tabular}
    \label{tab:compldm}
\end{table}

In Table\,\ref{tab:compldm} we show the RMSD values with respect to experimental total binding energies, calculated using the compressible LDM including pairing and shell effects. As shown, the RMSD tends to decrease with increasing $S_v$,
while it increases with increasing $L$. 
From these trends, one might expect the minimum RMSD to occur around $S_v >34$\,MeV, 
$50\,{\rm MeV}<L<70\,{\rm MeV}$. However, the actual location of the minimum may differ, as the saturation density 
  and incompressibility 
K are fixed in this table.
It should be noted that the RMSD values obtained from the compressible LDM are generally larger than those from the incompressible LDM. This is because in the compressible LDM, the central density and neutron skin are determined variationally to minimize the energy (see Eq.\,\eqref{eq:compldm}) for a given set of 
$(S_v, L, n_0, K)$, 
and these parameters have not yet been fully optimized.
The last column of Table\,\ref{tab:compldm} shows the number of isotopes for nickel ($Z=28$) nuclei. 
A smaller value of $S_v$ corresponds to a larger number of isotopes, whereas a smaller 
$L$ value tends to be associated with fewer isotopes. Interestingly, smaller RMSD values also correlate with a reduced number of isotopes.

In Figure \ref{fig:masstable} we show the binding energy differences (color bar) between experiment and the predictions from the optimized incompressible LDM, including shell effects using Eq.\eqref{eq:incompldm}. We also show as the pink symbol `$\boxtimes$' the stable isotopes out to the neutron dripline for the incompressible LDM. Finally, we show the neutron dripline predicted from the compressible LDM for a given $S_v$ and $L$. In the compressible LDM, we set $n_0=0.155\,\rm{fm}^{-3}$, $B=16.2\,\rm{MeV}$, and $K=240\,\rm{MeV}$. For each $(S_v,L)$, we find the pairing and shell coefficients that minimize the RMSD for nuclear binding energies. For light neutron-rich nuclei, the differences in the location of the dripline do not depend very sensitively on the value of the symmetry energy. On the other hand, in neutron-rich heavy nuclei, the neutron driplines are strongly affected by the value of the symmetry energy $S_v$ at fixed value of $L$. As the symmetry energy slope parameter $L$ increases, the dependence of the neutron dripline on the value of $S_v$ is reduced. In proton-rich nuclei, the isospin asymmetry is relatively small such that the nuclear symmetry energy effects are not apparent. The proton driplines predicted from the compressible LDM using Eq.\,\eqref{eq:compldm} with different symmetry energies are shown in Figure \ref{fig:masstable} and in all cases are very close to one another. Thus, the proton dripline is primarily determined by Coulomb repulsion.

In Figure \ref{fig:mtbsv} we show results identical to Figure \ref{fig:masstable}, except that now for the compressible LDM the symmetry energy $S_v$ is fixed and we allow the slope parameter $L$ to vary. Here we see that for light nuclei, a small value of $L = 30$\,MeV gives a much different neutron dripline than the two larger values $L = 50, 70$\,MeV. 
In contrast, for heavy neutron-rich nuclei, the location of the dripline has a strong dependence on $L$ within the range $30\,\text{MeV} < L < 70\,\text{MeV}$.


Note that including $L$ in the bulk nuclear matter energy, Eq.\ \eqref{eq:fbldm}, in the compressible LDM reduces the root-mean-square deviation for the binding energy of finite nuclei
when we allow $S_v$ and $L$ to vary. The domain for parameters $(B,n_0, K, S_v, L)$ 
is, however, restricted because the points for the least RMSD move to an unphysical region. 
In our work, we prepare for the domain, 
$n_0=(0.155,0.165)\,\rm{fm}^{-3}$, $B=(15.8, 16.5)$\,MeV, 
$K=(220,250)$\,MeV, 
$S_v=(28.5, 32.5)$\,MeV, 
and $L=(10,90)$\,MeV.
Compared to the results of Ref.\,\cite{Lattimer12a}, the lowest RMSD takes place at a somewhat different position, $(S_v,L) = (32.5,45.0)$\,MeV for the constrained band. 
\begin{figure}
    \centering
    \includegraphics[scale=0.8]{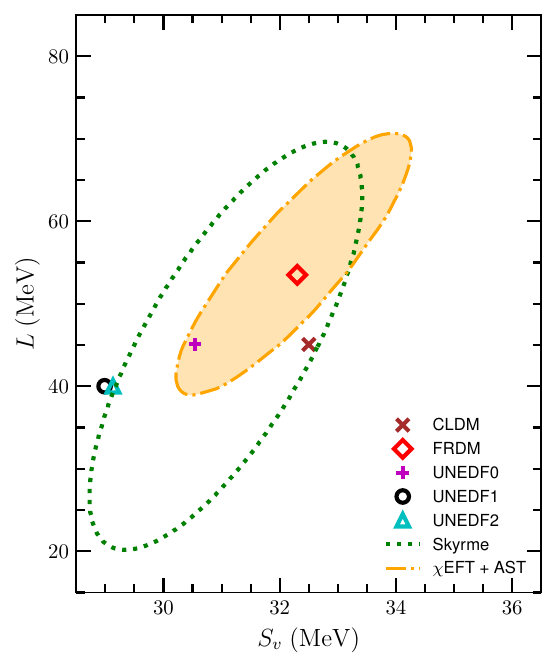}
    \caption{The symmetry energy parameters from various nuclear mass models. The ellipse constraints from
    the chiral effective calculations and astrophysical observations is added. }
    \label{fig:svlfinite}
\end{figure}
Figure \ref{fig:svlfinite} shows the symmetry energy parameters from the various nuclear mass models, FRDM\,\cite{frdm12}, 240 Skyrme averages\,\cite{dutra2012}, UNEDF0, UNEDF1, UNEDF2\,\cite{UNEDF0, UNEDF1, UNEDF2}, and the compressible liquid drop model of our work.
We have also added an ellipse contour that originates from $\chi$EFT constraints and astrophysical observations\,\cite{lim2024a}. 
As shown in the compilation of $S_v$ and $L$ in Ref.\ \cite{dutra2012}, the Skyrme force models exhibit considerable variation, rather than convergence, in their predicted values. The standard deviations are $16.23$\,MeV for $S_v$ and $60.63$\,MeV for $L$, 
with mean values of $29.32$\, MeV and $35.11$\,MeV, respectively.
However, by excluding Skyrme parameter sets that yield $S_v < 25$\,MeV or
$S_v > 40$\,MeV, and $L<0$\,MeV, which can be inferred from many-body calculations
with chiral interactions \cite{Hebeler10,gandolfi2012,gezerlis13,hagen14a,drischler16,Holt2016pjb}, 
the revised averages shift to $S_v=31.07$\,MeV 
and $L=44.90\,$MeV. These refined values fall within the uncertainty ellipse derived from chiral EFT predictions.
Fig. \ref{fig:svlfinite} shows the green contour corresponding to the $1 \sigma$ confidence interval for the $(S_v, L)$ values based on the 205 Skyrme interactions that satisfy the imposed constraints.
We included the $(S_v, L)$ values from the UNEDF models and FRDM, as these were developed for nuclear mass calculations. However, even among these nuclear mass models, the 
$(S_v, L)$ values do not exhibit convergence.
Compared to the ellipse contour constrained by $\chi$EFT, we see that most mass models, except FRDM and CLDM, prefer to have small $S_v$ and $L$. This suggests a slight tension between nuclear symmetry energy parameters determined from nuclear mass models and neutron star observations, which probe higher-density matter $n \sim 2-3n_0$.
\begin{figure}[t]
	\centering
	\includegraphics[scale=0.4]{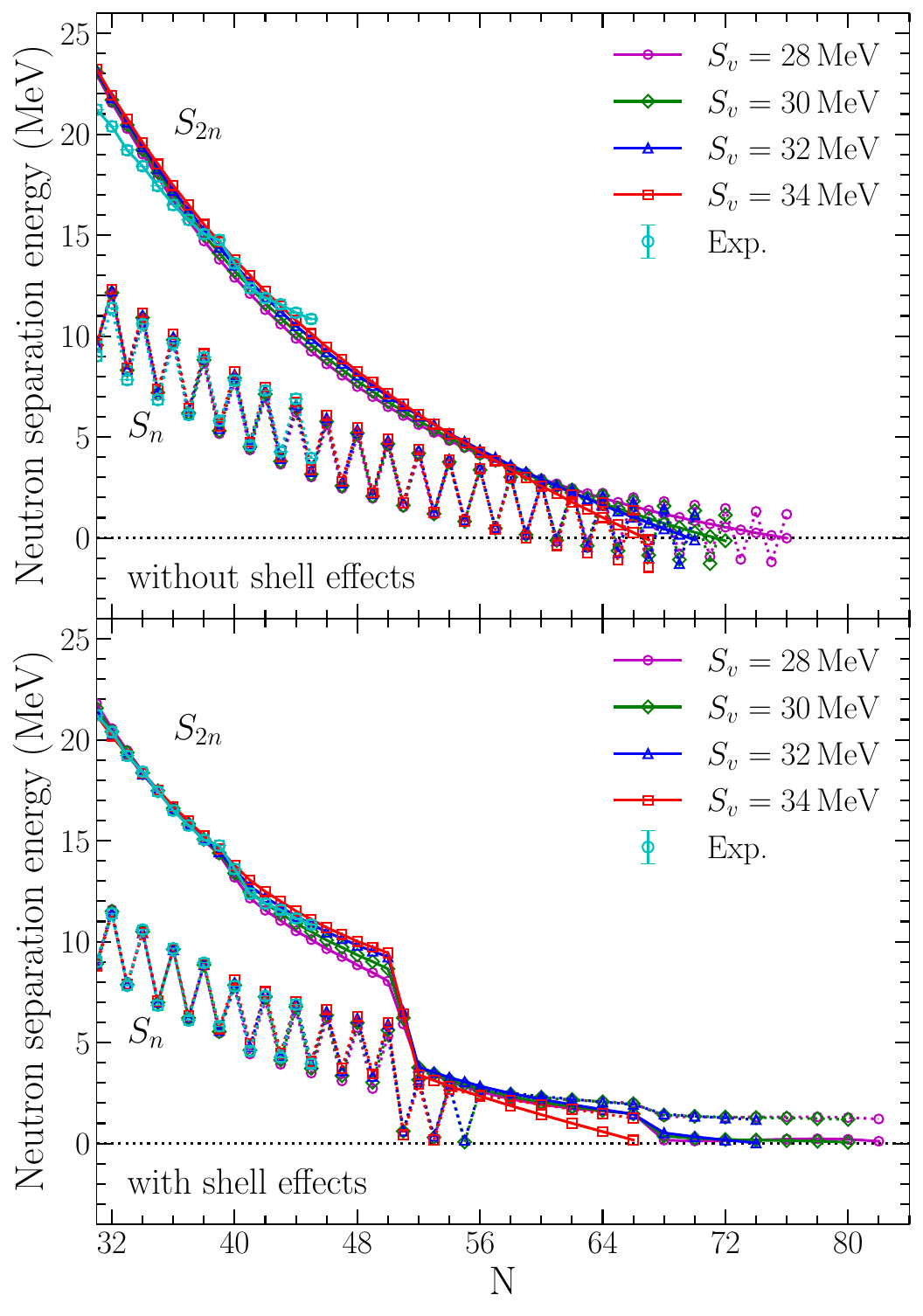}
	\caption{One-neutron and two-neutron separation energies for nickel isotopes in the compressible liquid drop model, neglecting shell effects (top panel) and including shell effects (bottom panel).
In each panel, the LDM models have varying $S_v$, with $L=50$ MeV held fixed. 
    }
	\label{fig:neutsepenergy}
\end{figure}

\begin{figure}
	\centering
	\includegraphics[scale=0.4]{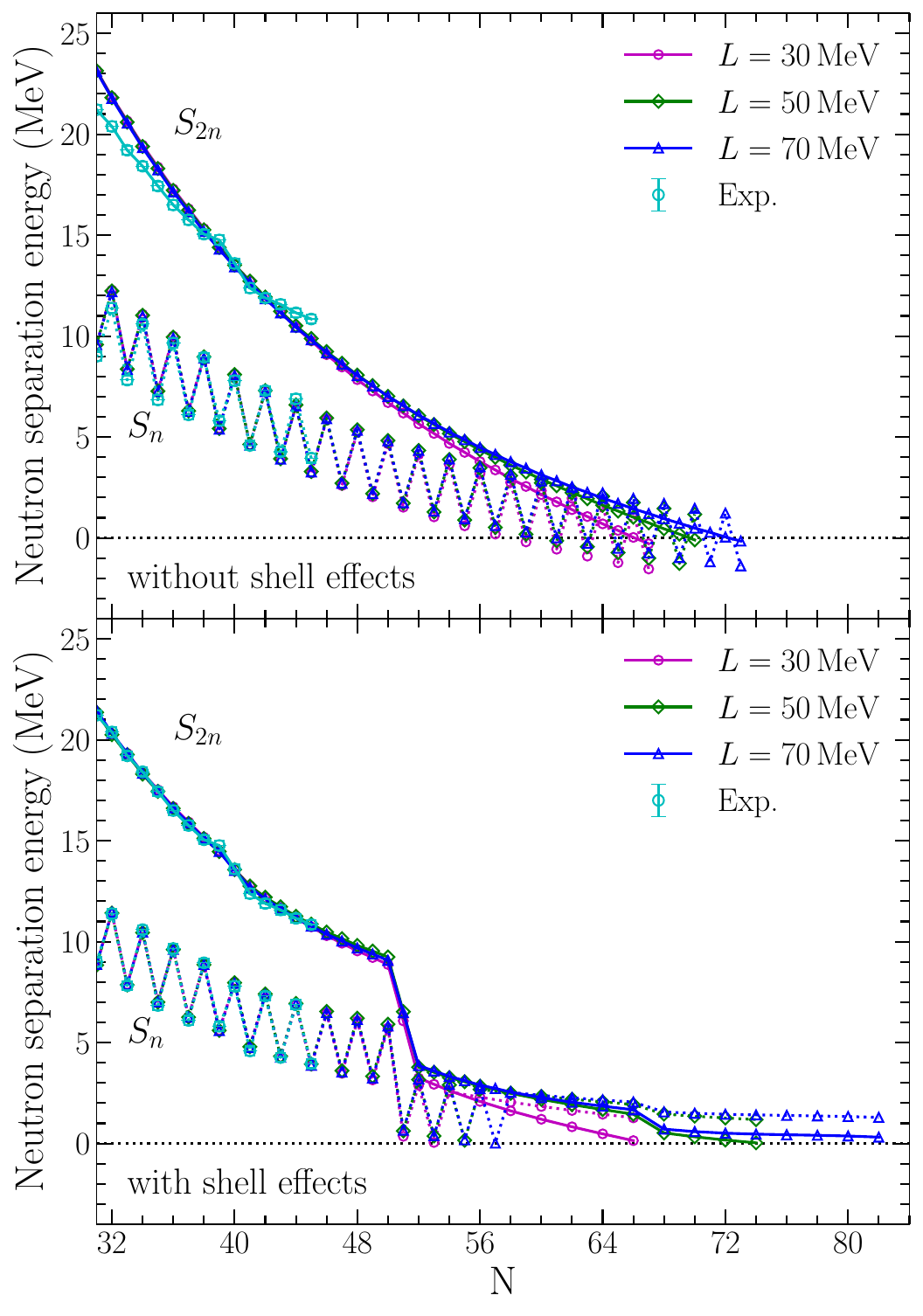}
	\caption{The same plot as in Fig. \ref{fig:neutsepenergy}, but with varying $L$ while keeping
    $S_v=32\,$MeV fixed.
    }
	\label{fig:neutsplv}
\end{figure}
\begin{figure}
	\centering
	\includegraphics[scale=0.53]{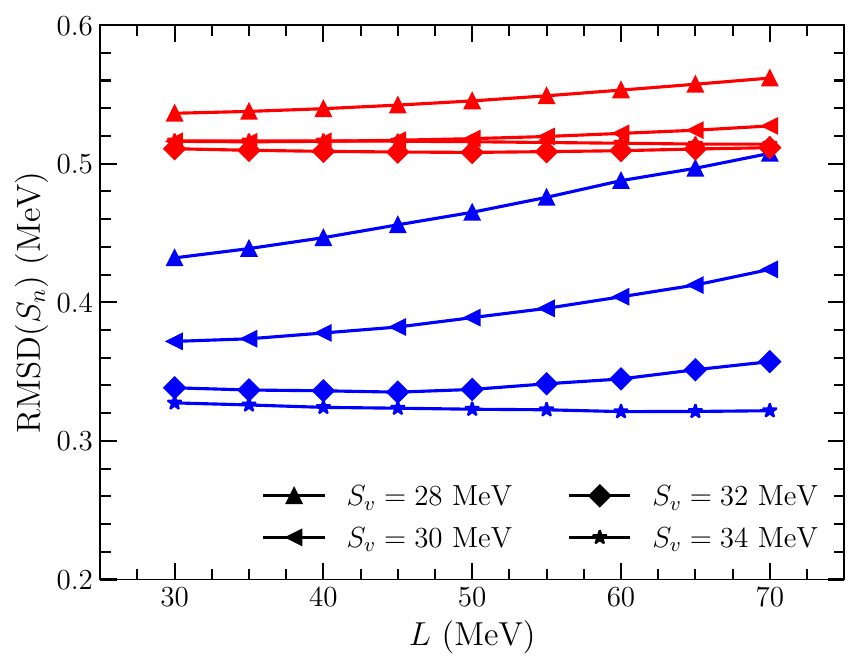}
	\caption{Root-mean-square deviations (RMSDs) of the one-neutron separation energy $S_n$ in the shell-model--corrected liquid-drop model (LDM) at fixed $S_v$. The red curves  
    represent the RMSD for all nuclei with measured masses ($Z \ge 8$); the blue curves indicate the RMSD for Ni isotopes only ($Z=28$).
}
	\label{fig:snsv32}
\end{figure}
\begin{figure}
	\centering
	\includegraphics[scale=0.53]{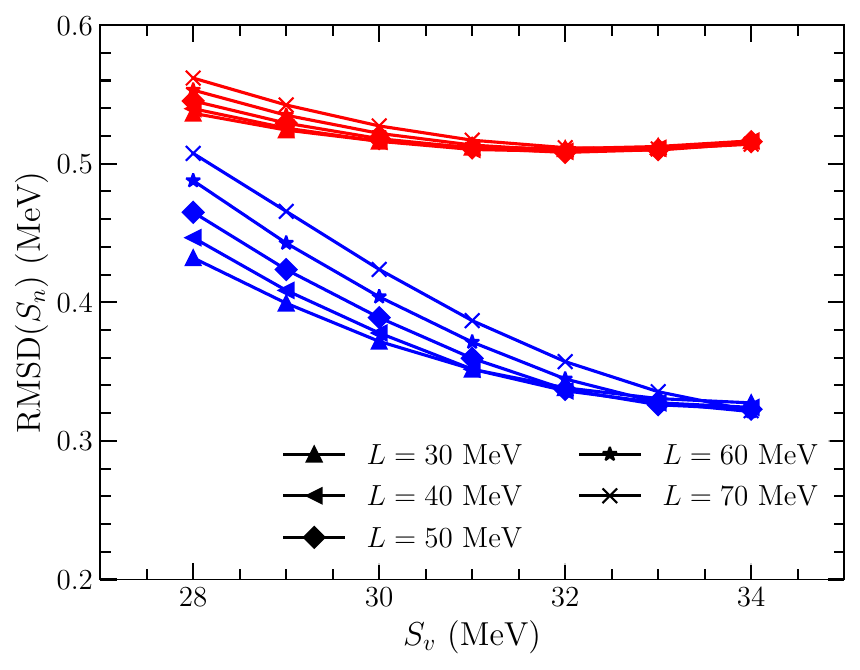}
	\caption{Same as Fig.\,\ref{fig:snsv32}, but with varying $S_v$ while keeping $L$ is fixed.}
	\label{fig:snlv50}
\end{figure}

From Figures \ref{fig:masstable} and \ref{fig:mtbsv}, one sees that the presence of shell closures strongly affect the location of the neutron dripline in certain regions of the nuclear chart. Therefore, in order to explore correlations among the symmetry energy, its slope parameter, and the neutron dripline, one should focus on specific regions of the nuclear chart away from shell closures. Figure\,\ref{fig:neutsepenergy} shows the one-neutron separation energies (dotted lines) and two-neutron separation energies (solid lines) for nickel isotopes 
from the compressible LDM for varying $S_v$ and fixed slope parameter $L=50\,\rm{MeV}$. We show results with shell corrections removed (top panel) and added (bottom panel). 
This even-odd staggering phenomenon originates from pairing correlations, which we employ in the compressible liquid drop model.
 Since the gap size is very similar for each model, $\Delta  \sim 12.0\,\mathrm{MeV}$, we expect that the one-neutron separation energy will have a similar behavior for each model. Compared to experiment, the one-neutron separation energies calculated within the compressible LDM show stronger variations with the mass number $A$. Overall, however, the experimental values are in general agreement with the LDM results. Compared to the one-neutron separation energies, the two-neutron separation energies calculated with the compressible LDM depend more sensitively on the symmetry energy $S_v$. When shell corrections are ignored, this gives rise to a large spread in the predicted location of the neutron dripline. When shell corrections are included, the neutron dripline in nickel is located in the range $N=66-82$, with a moderate dependence on the value of the symmetry energy.
The trend of the two-neutron separation energies computed with the compressible LDM including shell effects agrees well with experiment in the region where data exists. 
Figure \ref{fig:neutsplv} presents the same plot as Figure \ref{fig:neutsepenergy}, reaffirming that the compressible LDM with shell effects accurately describes both one-neutron and two-neutron separation energies, independent of the symmetry energy parameters. 
In this figure, we fix $S_v=32\,$MeV and vary $L=30$, $50$, and $70$\,MeV. 
This outcome naturally follows from the mass model, where shell effects contribute significantly to the nuclear mass. 
Figures~\ref{fig:snsv32} and \ref{fig:snlv50} show the RMSDs of the one-neutron separation energy $S_n$ at fixed $S_v$ and fixed $L$, respectively. 
The red curves indicate the RMSDs for all nuclei with measured masses ($Z \ge 8$), whereas the blue curves indicate the RMSDs for Ni isotopes only ($Z=28$). Although the numerical differences in the RMSDs are small and may be difficult to discern in Figures~\ref{fig:neutsepenergy} and \ref{fig:neutsplv}, they remain quantifiable: 
the RMSD is about $0.4\,\mathrm{MeV}$ for nickel isotopes and approximately 
$0.5\,\mathrm{MeV}$ when all experimentally measured nuclei are included. In the shell-corrected calculations, 
$S_n$ is nearly independent of $L$ for Ni isotopes, whereas the RMSD shows little dependence on $S_v$ when the full experimental dataset is considered.

We find again an important dependence of the neutron dripline with the symmetry energy slope parameter $L$.
It is worth noting that Figures 5 and 6 show the phenomenon $S_n > S_{2n}$ beyond $N > 56$. These numerical results arise from the interplay between pairing effects, proportional to $\Delta/\sqrt{A}$, and shell closures at $N_{\rm magic} \sim 8, 20, 28, 50$. The behavior of $S_{2n}$ beyond $N=56$ indicates that there is essentially no contribution from the pairing energy, since these nuclei are even–even, and no additional shell effects occur beyond $N=50$. Thus, in the absence of both pairing and shell contributions, $S_{2n}$ naturally tends toward zero.
In contrast, $S_{n}$ reflects the odd–even staggering, with even-$N$ nuclei bound and neighboring odd-$N$ nuclei unbound. This odd–even effect underlies the turnover observed in our results, where $S_{1n} > S_{2n}$.


\begin{figure*}
	\centering
	\includegraphics[scale=0.6]{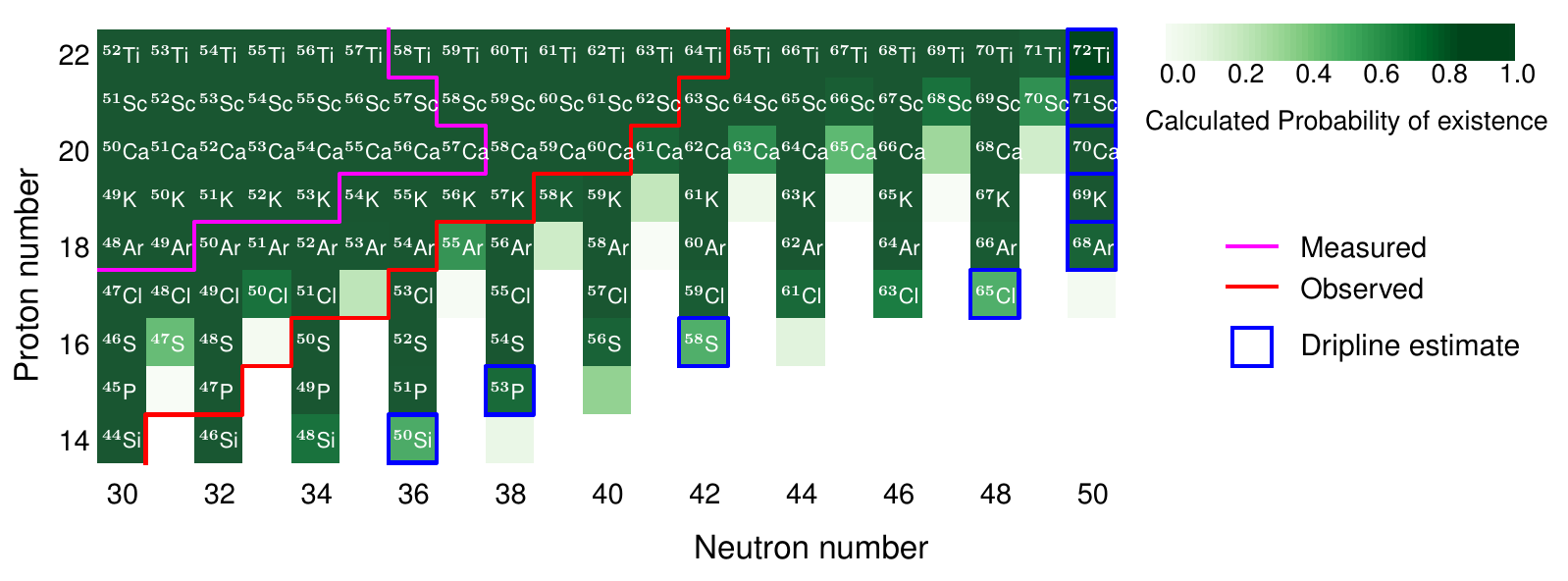}
	\caption{Existence probabilities for neutron-rich light nuclei out to the neutron dripline. The green color shading denotes the probability, and nuclei with less than 50\% probability of being bound are unlabeled in the figure. }
	\label{fig:lightchart}
\end{figure*}

Finally, since the location of the neutron dripline is sensitive to the choice of nuclear model, we investigate the robustness of our predictions by analyzing the results from 1000 energy density functionals, which are obtained from the posterior probability distribution obtained from combining nuclear experiment, nuclear theory, and astrophysical observations in Refs.\ \cite{Lim2018,lim19e}. We focus on the neutron driplines for the region $14\leq Z \leq 22$, which has recently been studied \cite{stroberg21} from ab initio nuclear many-body theory.
For the application to the LDM, we extract the nuclear matter properties for $B$, $n_0$, $S_v$, $L$, and $K$
from the EDFs. 
The missing terms for surface energy parameters in the EDFs
are obtained by fitting the experimental binding energies of 2028 finite nuclei. 
Based on this procedure, we construct 1,000 LDM mass tables.
Then, the probability for bound nuclei is taken as the fraction of mass tables 
in which a given isotope is bound. For example, if $^{58}\mathrm{S}$ has a probability $p=0.581$, 
this means that 581 out of the 1,000 LDM mass tables predict $^{58}\mathrm{S}$ to be bound.

Figure\,\ref{fig:lightchart} shows the probability that a particular nucleus will be bound within the compressible LDM. The label for each nucleus is present on the chart when the probability is greater than 0.5. 
The shell energy corrections are sufficiently strong that the neutron dripline from Ar to Ti is completed when $N=50$.
Recent state-of-the-art \emph{ab initio} calculations~\cite{stroberg21} predict that, in the Ar–Ti region, only the calcium isotopes have a neutron drip line beyond \(N=50\), with a probability between one-third and two-thirds.
Other light nuclei ($Z<26$) studied in Ref.\ \cite{stroberg21} have a neutron dripline below $N=50$. On the other hand, our results predict that the probabilities for the existence of $N=50$ are 0.956, 0.998, 1.0, 1.0, and 1.0 for $^{18}$Ar, $^{19}$K, $^{20}$Ca, $^{21}$Sc, and $^{22}$Ti respectively. The density functional approach \cite{neufcourt19}, however, predicts that both Sc and Ti have more than 50 neutrons at their driplines.

\begin{figure*}[t]
\centering
\includegraphics[scale=0.74]{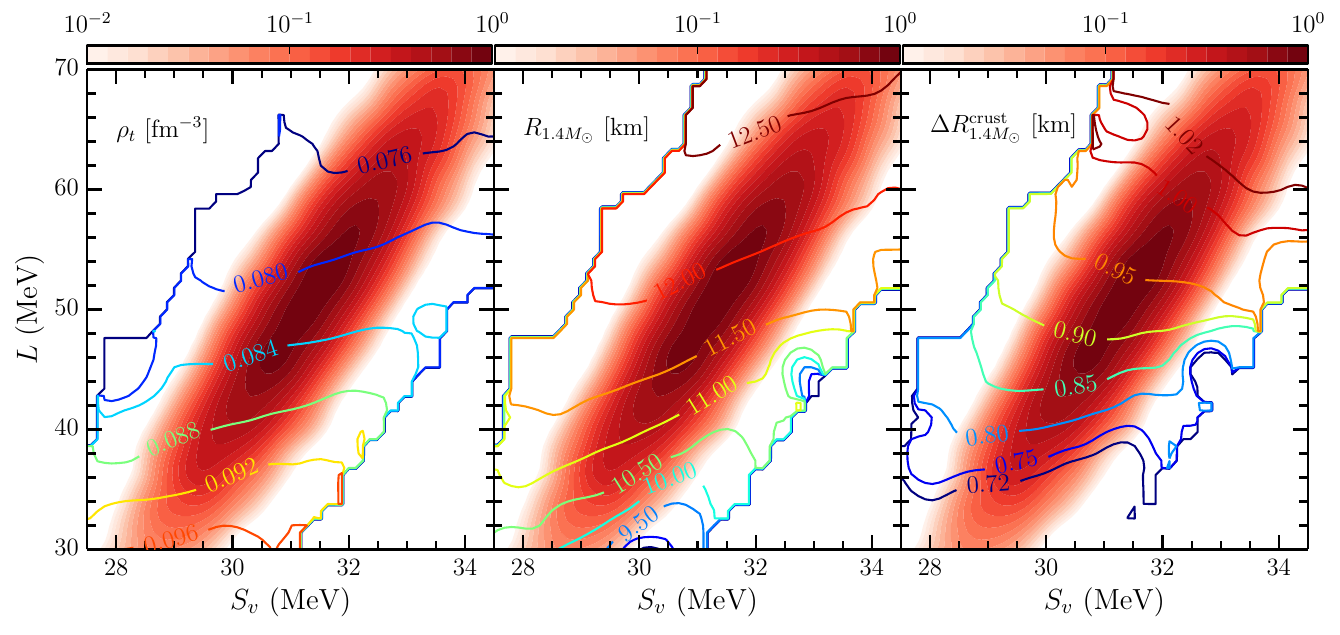}
\caption{(Color online) Neutron star core-crust transition density (left), radius (middle), and crust thickness (right) for a 1.4~$\msun$ neutron star and given values of $S_v$ and $L$.
The probability distribution of the red ellipses was generated using 1,000 EDFs.
}

\label{fig:crust}
\end{figure*}

\section{LIQUID DROP MODEL AND THE NEUTRON STAR CRUST}
\label{results}
In this section we explore correlations between LDM parameters, neutron star crustal properties, and connections with finite nuclei.
The connections between the crust-core transition density, the crustal thickness, the crustal composition, and the neutron star radius to the symmetry energy $S_v$ and its slope parameter $L$
are investigated.
Finally, we show that the last bound nuclei in selected isotopes are well correlated with neutron star radii.

\subsection{Nuclear symmetry energy and core-crust boundaries}
The core-crust boundary in neutron stars can be found by increasing the density of beta-equilibrium matter and comparing the energy density of the inhomogeneous (crust) phase to the uniform nuclear matter phase. An alternative approach starts by decreasing the density of the uniform nuclear matter phase and analyzing the appearance of thermodynamic instabilities that indicate the onset of the inhomogeneous phase ~\cite{baym71,pethick95,kubis07,lattimer07rep,hebeler13}. In the present work, we employ the first method that considers the total energy of the system. The phase transition from inhomogeneous nuclear matter to uniform nuclear matter occurs through a competition between the Coulomb energy and the nuclear surface energy. Hence, the correct formalism in the inner crust of 
neutron stars as well as the accurate nuclear energy density functionals
are necessary to determine core-crust transition densities and the crust thickness.

In order to relate the symmetry energy $S_v$ and its density derivative $L$ to the core-crust transition density and the crust thickness, we have employed the energy functional,
\begin{align}
\mathcal{E}_B(n,x)
& =  \frac{1}{2m}\tau_n + \frac{1}{2m}\tau_p \label{eq:eden} \\
&+  (1-2x)^2 f_n(n) + \left[1 -(1-2x)^2\right] f_s(n)\,, \nonumber
\end{align}
where $n$ is the baryon number density, $x$ is the proton fraction, $\tau_n$ and $\tau_p$ are the kinetic energy densities for neutrons and protons, respectively, and $f_n$ and $f_s$ correspond to the pure neutron matter and symmetric nuclear matter potential energy functions for a given density for which we use the expansion:
\begin{equation}
f_s(n) = \sum_{i=0}^{3} a_i n^{(2+i/3)}\,
,\quad
f_n(n) = \sum_{i=0}^{3} b_i n^{(2+i/3)}\,.
\end{equation}
For symmetric matter, the $a_i$'s are determined from nuclear matter equation of state empirical parameters, such as $n_0$, $B$, $K$, and $Q$~\cite{Lim2018}. In this work, we fit the $b_i$'s to neutron matter calculations \cite{PhysRevC.87.014338} based on five chiral EFT interactions from the literature. 
The core-crust boundary is then found by comparing the energy difference between inhomogeneous nuclear matter and uniform nuclear matter composed of neutrons, protons, and electrons, similar to the liquid drop model technique employed in previous work \cite{lim2017c} 
%

In Figure \ref{fig:crust} we show the core-crust transition density (left panel), the radius (middle panel), and the crust thickness (right panel) for 1.4~$\msun$ neutron stars. The red ellipse region denotes the $S_v - L$ probability distribution for the 1,000 energy density functionals used in this work that were originally generated from a Bayesian statistical analysis of the nuclear equation of state constrained by chiral effective field theory and empirical properties of nuclei \cite{Lim2018}. 

In general, the transition density increases slowly as $S_v$ increases. On the other hand, the transition density decreases more strongly as $L$ increases because the pressure of uniform nuclear matter increases linearly as $p(n) \simeq \frac{L}{3}n$ and therefore the phase transition takes place at lower densities. The central ellipse from the energy density functionals indicates that the core-crust transition density lies in the range $\rho_t=0.076~\text{fm}^{-3} < \rho_t < \rho_t=0.096~\text{fm}^{-3}$. Our results agree with those of Ref.\ \cite{hebeler13}, where the thermodynamic instability condition was used to find the transition density in terms of the Coulomb energy and the density gradient terms $Q_{nn}$ and $Q_{np}$, while here we used the liquid drop model to compare the energy density between uniform nuclear matter and inhomogeneous nuclear matter.
As you can see the first panel in Fig.\,\ref{fig:crust}, there exists correlation between $\rho_t$ and $S_v$ and anti-correlation between $\rho_t$ and $L$.
This suggests the linear fitting for $\rho_t$ of $S_v$ and $L$,
\begin{equation}\label{eq:rhotsvl}
    \rho_t = \rho_{t_0} + \eta S_v + \zeta L .
\end{equation}
Table\,\ref{tb:rhotsvl} shows the numerical values for parameters in Eq.\,\eqref{eq:rhotsvl}.
The negative sign for $\zeta$, as expected, indicates that there is a anti-correlation 
between $\rho_t$ and $L$ which was observed in Ref.\,\cite{lim2019moi}.
It is interesting that $S_v$ has the positive correlation however $L$ is anti-correlated
with $\rho_t$ even though $S_v$ and $L$ are strongly correlated as can be seen
in Fig.\,\ref{fig:crust}.
This is because the pressure at the core–crust boundary is largely determined by the relation $p_t \sim \frac{L}{3}\rho)_t$.
Since the proton fraction near the transition density is only about $3\%$, the pressure is dominated by neutrons.
\begin{table}[t]
	\caption{Numerical values for parameters in Eq.~\eqref{eq:rhotsvl}.}
	\begin{tabular}{cc}
		\hline
		\hline
		\\[-2.8ex]
		$\rho_{t_0}$ (fm$^{-3}$)         & $7.014\times 10^{-2} \pm 2.535\times 10^{-3}$ \\
		\hline
		\\[-2.8ex]
		$\eta$ (MeV$^{-1}$~fm$^{-3}$)  &  $1.640\times 10^{-3} \pm 8.108\times 10^{-5}$ \\
		\hline
		\\[-2.8ex]
		$\zeta$ (MeV$^{-1}$~fm$^{-3}$)   &  $-7.734\times 10^{-4} \pm 5.587\times 10^{-5}$ \\
		\\[-2.8ex]
		\hline
	\end{tabular}\label{tb:rhotsvl}
\end{table}

The radius of a $1.4\,\msun$ neutron star increases as $S_v$ decreases and $L$ is kept fixed. If the symmetry energy is large, then beta-equilibrium nuclear matter will tend to have a larger proton fraction, which will result in a reduced overall pressure and neutron star radius. On the other hand, increasing $L$ at fixed symmetry energy $S_v$ increases the pressure and the overall neutron star radius. The crust thickness of a $1.4~\msun$ neutron star increases as $L$ increases. The crust thickness is defined as the difference between the whole radius and core radius where the core-crust transition happens. Even if the transition density decreases as $L$ increases, the total radius increases faster than the core radius, and therefore the crust thickness increases in general. The correlation between $S_v$ and the crust thickness, however, is very weak due to competing effects in the energy density functional for $S_v$. The crust thickness from the present analysis is in the range $0.7\,\text{km} < \Delta R < 1.0\,\text{km}$.

\begin{figure}
	\centering
	\includegraphics[scale=0.58]{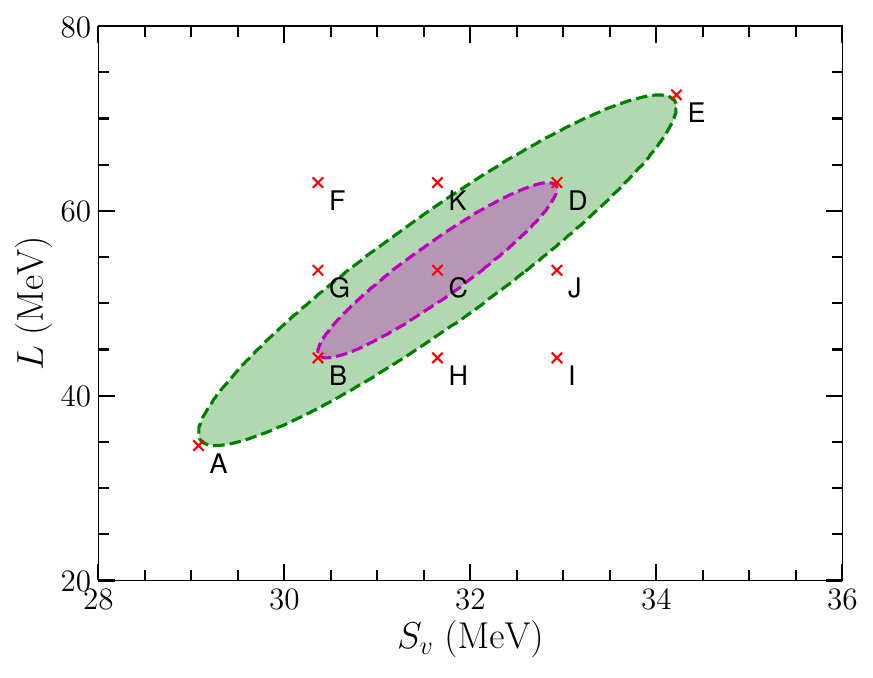}
	\caption{$S_v$ and $L$ correlation contour plot with $2\sigma$ uncertainties ($1\sigma$ inside). 
		The symbol `$\mathbf{x}$' exhibits the representative ($S_v$, $L$) for the purpose of comparison
	between PNM calculation from the chiral perturbation theory and  energy density functionals.}
	\label{fig:svlcont}
\end{figure}

\begin{figure}[t]
	\centering
	\includegraphics[scale=0.58]{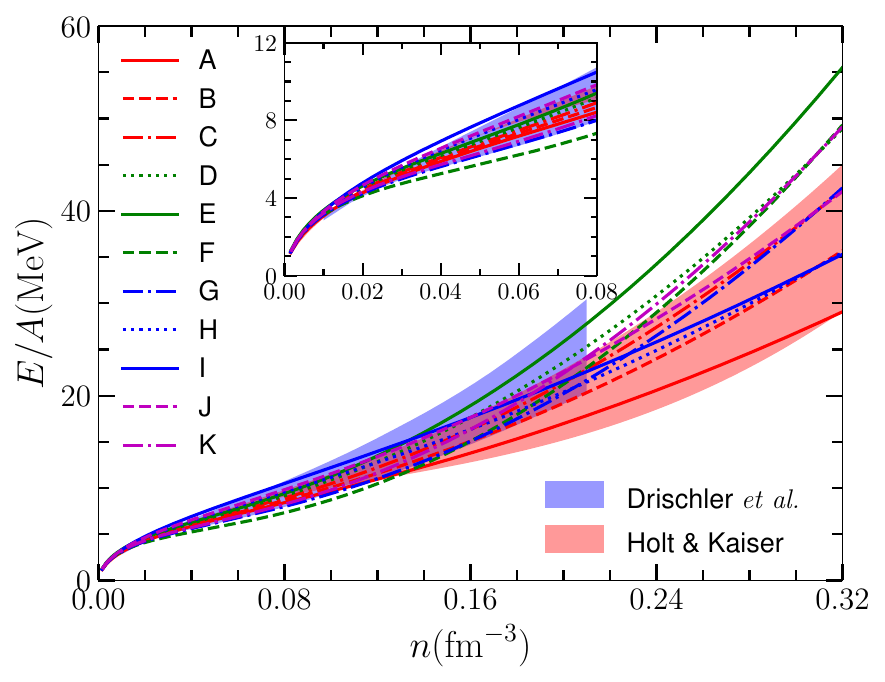}
	\caption{Pure neutron matter equation of state uncertainty bands from chiral effective field theory and from the energy density functionals with corresponding ($S_v, L$) parameters as in Fig.\,\ref{fig:svlcont}.}
	\label{fig:svlpnmbandsub}
\end{figure}

In order to disentangle specific effects from $S_v$ and $L$, we orient our discussion according to Figure\,\ref{fig:svlcont}, which shows the correlated confidence interval between $S_v$ and $L$ predicted through a combination of chiral EFT equation of state calculations and empirical constraints developed in previous work \cite{Lim2018, lim19e}. The inner (outer) ellipse corresponds to the $1\sigma$ ($2\sigma$) range of credibility. We also denote several points from `A' to `K' to connect the ($S_v$, $L$) values with the energy density functionals and equation of state of pure neutron matter (PNM). Figure\,\ref{fig:svlpnmbandsub} shows the energy per baryon of pure neutron matter as a function of density obtained from energy density functionals that use the $(S_v,L)$ parameter sets as in Fig.\ \ref{fig:svlcont}. For comparison, we have also included two theoretical calculations by Drischler \textit{et al.} \cite{drischler16} (blue band) and Holt \& Kaiser \cite{Holt2016pjb} (red band). In general, we observe that model `A' (lowest values of ($S_v,L$)) and `E' (highest values of ($S_v,L$)) give rise to the softest and stiffest equations of state, respectively. However, the stiffness of the equations of state across all densities has a more complicated dependence. For instance, point `F' (green dashed line) has a low value of $S_v$ and a high value of $L$. This produces an equation of state curve with the smallest energy density at low densities but one of the highest energy densities at high density, relative to the other models. Correspondingly, model `I' (solid blue line) has a high value of $S_v$ and a low value of $L$. This produces an equation of state curve with the highest energy density at low densities but one of the smallest energy densities at high density. In general, the models with large $L$ give rise to equations of state with a large energy density at the highest densities, but they can have widely varying energies at low density.

\begin{figure}[b]
	\centering
	\includegraphics[scale=0.58]{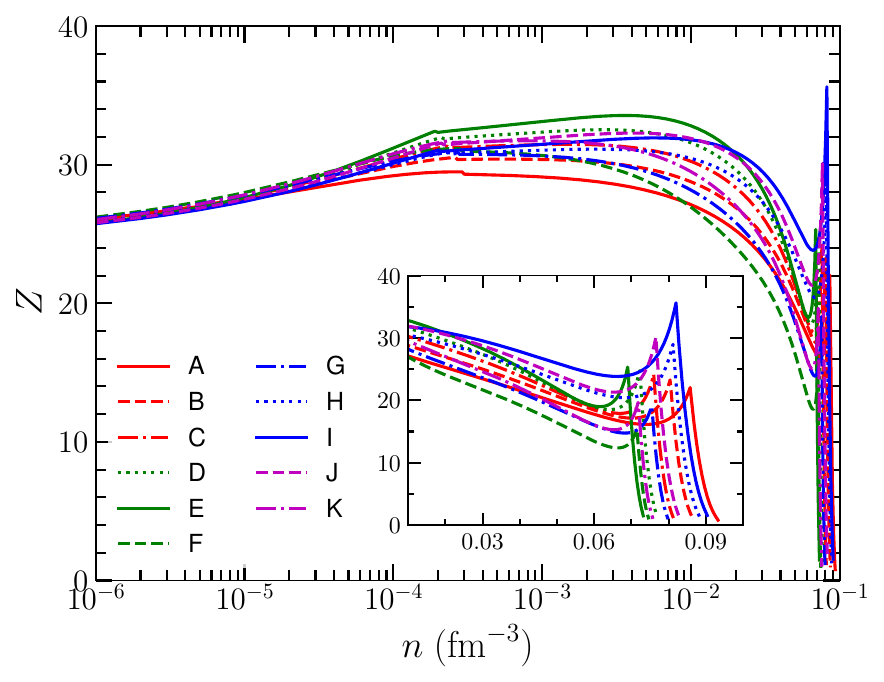}
	\caption{Atomic number of heavy nuclei in the crusts of neutron stars for varying values of the symmetry energy $S_v$ and slope parameter $L$ (see Figure \ref{fig:svlcont} for details).}
	\label{fig:svlcrustazsub}
\end{figure}

Figure\,\ref{fig:svlcrustazsub} shows the atomic number in the inner and outer crusts of neutron stars. Most of the empirical models from `A' to `K' studied in this work are expected to give similar atomic numbers in the outer crusts of neutron stars. The deviations start to become large above a baryon number density of $10^{-4}$\,fm$^{-3}$. In addition, one can observe a small discontinuity in the slope of the atomic number with respect to baryon number density around $n \sim 2\times 10^{-4}$\,fm$^{-3}$. This is the region where the neutrons first begin to drip out of heavy nuclei. The inset of Fig.\ \ref{fig:svlcrustazsub} shows the atomic number in the baryon number density range between 0.01 and 0.1\,fm$^{-3}$. The local maximum for each curve corresponds to the density where the bubble phase appears. In terms of the volume fraction of dense matter to the total baryon number density, it is the density where $u = 0.5$, that is, the slab phase. As the total baryon number density increases beyond this point, the atomic number decreases until the inhomogenous phase turns into the uniform nuclear matter phase. Model `I' (high $S_v$ and low $L$) in general predicts the largest atomic numbers in neutron star crusts, where as Model `F' (low $S_v$ and high $L$) in general predicts the smallest atomic numbers. This can be understood from the fact that model `I' shows the largest energy per baryon in pure neutron matter at low densities, as can be seen in Figure \ref{fig:svlpnmbandsub}. In the same manner, model `F' shows the lowest energy per baryon in pure neutron matter at low densities, and therefore this model gives the lowest atomic number in the sub nuclear density range. These results indicate that PNM calculations can give important insights into the composition of neutron star inner crusts.
The correlation between pure neutron matter and energy density functionals, particularly those based on the Skyrme Hartree-Fock framework, has been extensively studied by many authors\,\cite{Chabanat1997,Stone2003,Goriely2005}. These studies consistently show that neutron-rich nuclei, especially those near the neutron dripline, are strongly influenced by the properties of neutron matter. Therefore, it is a natural conclusion that accurate descriptions of neutron matter should be taken into account when constructing energy density functionals.

\begin{figure}[t]
	\centering
	\includegraphics[scale=0.58]{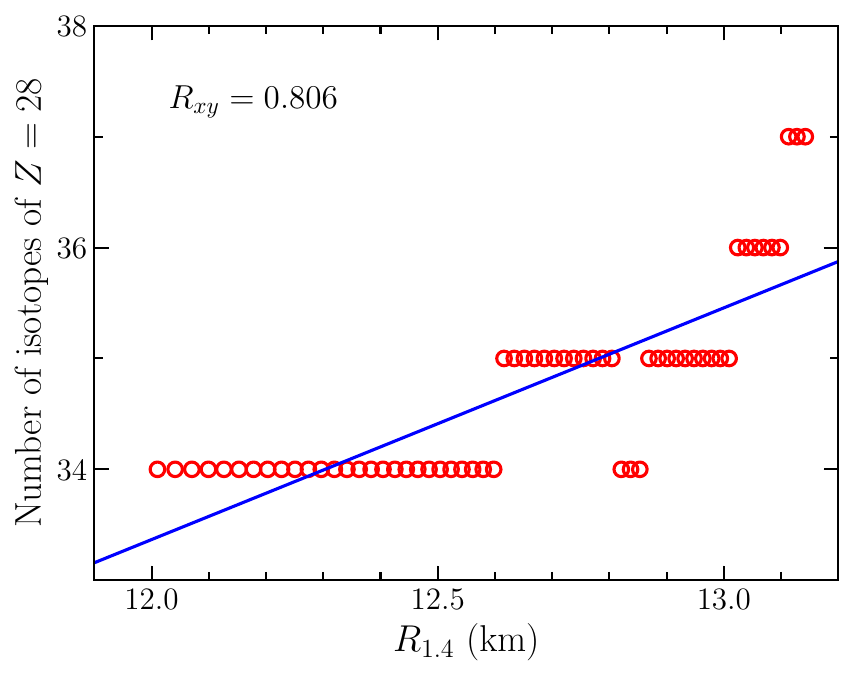} \\
 \includegraphics[scale=0.58]{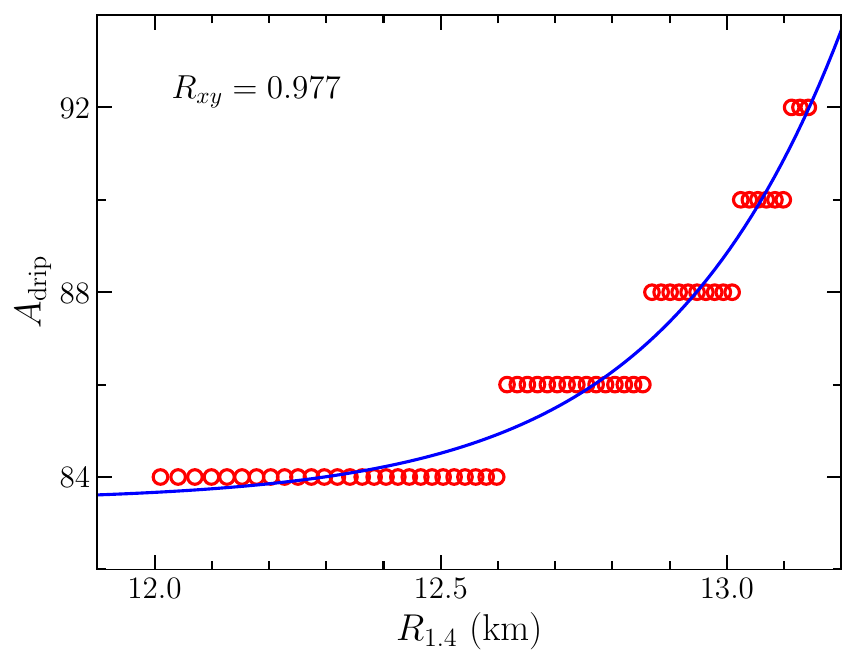}
	\caption{(Top) Correlation between the number of bound nickel isotopes and the radius of a $1.4\,\msun$ neutron star. (Bottom) Correlation between the mass number of the nickel neutron drip line and the radius of a $1.4\,\msun$ neutron star. 
 }
	\label{fig:r14z28}
\end{figure}

\subsection{Neutron drip and radius of neutron stars}

Finally, in upper panel of Figure \ref{fig:r14z28} we show the correlation between the number of nickel isotopes 
($Z=28$) and the radius $R_{1.4}$ of a $1.4\msun$ neutron star for
a given equation of state. 
In the lower panel of Figure \ref{fig:r14z28} the correlation between the mass number of last bound
nickel isotopes and the radius of a $1.4\msun$ neutron star are shown for
$61$ equations of state. 
We observe a strong correlation ($R_{xy}=0.977$) between the nickel dripline mass number and $R_{1.4}$
when we use the fitting function of $A_{\mathrm{drip}} = a + b \left(\frac{R_{1.4}}{12\,\rm{km}} \right)^{\alpha}$, where $\alpha=40$.
Our algebraic shell correction formula, Eq.\ \eqref{shellcor}, predicts that the first long plateau ($\#~\rm{iso.} = 34$) in Figure \ref{fig:r14z28} appears when $L=20$ MeV and lasts until $L=46$ MeV. The last bound nucleus corresponds $Z=28$ and $N=56$. 
When defining the shell corrections in the LDM, the reference magic numbers are taken to be $8$, $20$, $28$, $50$, and $82$. 
Since the mass number of the last-bound nucleus and $R_{1.4}$ both increase as $L$ increases~\cite{lim2024a}, it is natural to predict that the number of isotopes has a strong correlation with $R_{1.4}$.

\section{Conclusions}
\label{sec:con}

We have studied the role of the nuclear symmetry energy in the determination of neutron driplines and the consequences for neutron star radii and composition of the crust. 
We utilized the liquid drop model with pairing effects and algebraic nuclear shell corrections. We have shown that for selected isotopic chains, the location of the neutron dripline is particularly sensitive to the symmetry energy and its slope parameter. On the other hand, the proton dripline is not very sensitive to the symmetry energy because the Coulomb energy is the dominant factor to determine the bound states of proton-rich nuclei.

We have found that the core-crust transition density is positively correlated with $S_v$ and anti-correlated with $L$.
We also confirm the well known  correlation between the symmetry energy slope parameter $L$ and the radius of a $1.4\,\msun$ star. Furthermore, as $L$ increases, the crust thickness increases because large values of $L$ increase the radius of neutron stars as predicted. We have shown that the atomic number of nuclei in the crust of neutron stars is highly correlated with the pure neutron matter equation of state. In general, a large energy per particle of neutron matter, $E/A$, is associated with the presence of nuclei with large atomic numbers $Z$ in the crusts of neutron stars. Thus, theoretical neutron matter calculations are an important benchmark to study heavy nuclei in the crust of neutron stars. We have found that the driplines of selected isotopes may be strongly correlated with neutron star bulk properties. In particular, radius of $1.4\,\msun$ neutron stars was shown to vary sensitively with the mass number of the nickel dripline.

\acknowledgments
Y. Lim was supported by the National
Research Foundation of Korea(NRF) grant funded by the
Korea government(MSIT)(No. RS-2024-00457037) and  by the Yonsei University Research
Fund of 2024-22-0121. 
Y. Lim was also supported by Global - Learning \& Academic research institution for Master’s·PhD students, 
and Postdocs(LAMP) Program of the National Research Foundation of Korea(NRF) grant funded by the Ministry of Education(No.  RS-2024-00442483).
J.~W. Holt was supported in part by the National Science Foundation under grant No.\ PHY2209318.


\begin{thebibliography}{86}%
	\makeatletter
	\providecommand \@ifxundefined [1]{%
		\@ifx{#1\undefined}
	}%
	\providecommand \@ifnum [1]{%
		\ifnum #1\expandafter \@firstoftwo
		\else \expandafter \@secondoftwo
		\fi
	}%
	\providecommand \@ifx [1]{%
		\ifx #1\expandafter \@firstoftwo
		\else \expandafter \@secondoftwo
		\fi
	}%
	\providecommand \natexlab [1]{#1}%
	\providecommand \enquote  [1]{``#1''}%
	\providecommand \bibnamefont  [1]{#1}%
	\providecommand \bibfnamefont [1]{#1}%
	\providecommand \citenamefont [1]{#1}%
	\providecommand \href@noop [0]{\@secondoftwo}%
	\providecommand \href [0]{\begingroup \@sanitize@url \@href}%
	\providecommand \@href[1]{\@@startlink{#1}\@@href}%
	\providecommand \@@href[1]{\endgroup#1\@@endlink}%
	\providecommand \@sanitize@url [0]{\catcode `\\12\catcode `\$12\catcode
		`\&12\catcode `\#12\catcode `\^12\catcode `\_12\catcode `\%12\relax}%
	\providecommand \@@startlink[1]{}%
	\providecommand \@@endlink[0]{}%
	\providecommand \url  [0]{\begingroup\@sanitize@url \@url }%
	\providecommand \@url [1]{\endgroup\@href {#1}{\urlprefix }}%
	\providecommand \urlprefix  [0]{URL }%
	\providecommand \Eprint [0]{\href }%
	\providecommand \doibase [0]{http://dx.doi.org/}%
	\providecommand \selectlanguage [0]{\@gobble}%
	\providecommand \bibinfo  [0]{\@secondoftwo}%
	\providecommand \bibfield  [0]{\@secondoftwo}%
	\providecommand \translation [1]{[#1]}%
	\providecommand \BibitemOpen [0]{}%
	\providecommand \bibitemStop [0]{}%
	\providecommand \bibitemNoStop [0]{.\EOS\space}%
	\providecommand \EOS [0]{\spacefactor3000\relax}%
	\providecommand \BibitemShut  [1]{\csname bibitem#1\endcsname}%
	\let\auto@bib@innerbib\@empty
	\bibitem [{\citenamefont {Myers}\ and\ \citenamefont
		{Swiatecki}(1969)}]{Myers69}%
	\BibitemOpen
	\bibfield  {author} {\bibinfo {author} {\bibfnamefont {W.}~\bibnamefont
			{Myers}}\ and\ \bibinfo {author} {\bibfnamefont {W.}~\bibnamefont
			{Swiatecki}},\ }\href@noop {} {\bibfield  {journal} {\bibinfo  {journal}
			{Ann. Phys. (N.Y.) 55: 395-505.}\ } (\bibinfo {year} {1969})}\BibitemShut
	{NoStop}%
	\bibitem [{\citenamefont {Lattimer}\ and\ \citenamefont
		{Lim}(2013)}]{Lattimer12a}%
	\BibitemOpen
	\bibfield  {author} {\bibinfo {author} {\bibfnamefont {J.~M.}\ \bibnamefont
			{Lattimer}}\ and\ \bibinfo {author} {\bibfnamefont {Y.}~\bibnamefont {Lim}},\
	}\href@noop {} {\bibfield  {journal} {\bibinfo  {journal} {Astrophys. J.}\
		}\textbf {\bibinfo {volume} {771}},\ \bibinfo {pages} {51} (\bibinfo {year}
		{2013})}\BibitemShut {NoStop}%
	\bibitem [{\citenamefont {Pearson}\ \emph {et~al.}(2014)\citenamefont
		{Pearson}, \citenamefont {Chamel}, \citenamefont {Fantina},\ and\
		\citenamefont {Goriely}}]{Pearson2014}%
	\BibitemOpen
	\bibfield  {author} {\bibinfo {author} {\bibfnamefont {J.~M.}\ \bibnamefont
			{Pearson}}, \bibinfo {author} {\bibfnamefont {N.}~\bibnamefont {Chamel}},
		\bibinfo {author} {\bibfnamefont {A.~F.}\ \bibnamefont {Fantina}}, \ and\
		\bibinfo {author} {\bibfnamefont {S.}~\bibnamefont {Goriely}},\ }\href@noop
	{} {\bibfield  {journal} {\bibinfo  {journal} {Eur. Phys. J. A}\ }\textbf
		{\bibinfo {volume} {50}},\ \bibinfo {pages} {43} (\bibinfo {year}
		{2014})}\BibitemShut {NoStop}%
	\bibitem [{\citenamefont {Chen}\ \emph {et~al.}(2005)\citenamefont {Chen},
		\citenamefont {Ko},\ and\ \citenamefont {Li}}]{chen05}%
	\BibitemOpen
	\bibfield  {author} {\bibinfo {author} {\bibfnamefont {L.-W.}\ \bibnamefont
			{Chen}}, \bibinfo {author} {\bibfnamefont {C.~M.}\ \bibnamefont {Ko}}, \ and\
		\bibinfo {author} {\bibfnamefont {B.-A.}\ \bibnamefont {Li}},\ }\href@noop {}
	{\bibfield  {journal} {\bibinfo  {journal} {Phys. Rev. C}\ }\textbf {\bibinfo
			{volume} {72}},\ \bibinfo {pages} {064309} (\bibinfo {year}
		{2005})}\BibitemShut {NoStop}%
	\bibitem [{\citenamefont {Centelles}\ \emph {et~al.}(2009)\citenamefont
		{Centelles}, \citenamefont {Roca-Maza}, \citenamefont {Vi\~nas},\ and\
		\citenamefont {Warda}}]{centelles09}%
	\BibitemOpen
	\bibfield  {author} {\bibinfo {author} {\bibfnamefont {M.}~\bibnamefont
			{Centelles}}, \bibinfo {author} {\bibfnamefont {X.}~\bibnamefont
			{Roca-Maza}}, \bibinfo {author} {\bibfnamefont {X.}~\bibnamefont {Vi\~nas}},
		\ and\ \bibinfo {author} {\bibfnamefont {M.}~\bibnamefont {Warda}},\
	}\href@noop {} {\bibfield  {journal} {\bibinfo  {journal} {Phys. Rev. Lett.}\
		}\textbf {\bibinfo {volume} {102}},\ \bibinfo {pages} {122502} (\bibinfo
		{year} {2009})}\BibitemShut {NoStop}%
	\bibitem [{\citenamefont {Roca-Maza}\ \emph {et~al.}(2011)\citenamefont
		{Roca-Maza}, \citenamefont {Centelles}, \citenamefont {Vi\~nas},\ and\
		\citenamefont {Warda}}]{Roca2011}%
	\BibitemOpen
	\bibfield  {author} {\bibinfo {author} {\bibfnamefont {X.}~\bibnamefont
			{Roca-Maza}}, \bibinfo {author} {\bibfnamefont {M.}~\bibnamefont
			{Centelles}}, \bibinfo {author} {\bibfnamefont {X.}~\bibnamefont {Vi\~nas}},
		\ and\ \bibinfo {author} {\bibfnamefont {M.}~\bibnamefont {Warda}},\
	}\href@noop {} {\bibfield  {journal} {\bibinfo  {journal} {Phys. Rev. Lett.}\
		}\textbf {\bibinfo {volume} {106}},\ \bibinfo {pages} {252501} (\bibinfo
		{year} {2011})}\BibitemShut {NoStop}%
	\bibitem [{\citenamefont {Oyamatsu}\ \emph {et~al.}(2010)\citenamefont
		{Oyamatsu}, \citenamefont {Iida},\ and\ \citenamefont {Koura}}]{oyamatsu10}%
	\BibitemOpen
	\bibfield  {author} {\bibinfo {author} {\bibfnamefont {K.}~\bibnamefont
			{Oyamatsu}}, \bibinfo {author} {\bibfnamefont {K.}~\bibnamefont {Iida}}, \
		and\ \bibinfo {author} {\bibfnamefont {H.}~\bibnamefont {Koura}},\
	}\href@noop {} {\bibfield  {journal} {\bibinfo  {journal} {Phys. Rev. C}\
		}\textbf {\bibinfo {volume} {82}},\ \bibinfo {pages} {027301} (\bibinfo
		{year} {2010})}\BibitemShut {NoStop}%
	\bibitem [{\citenamefont {Wang}\ and\ \citenamefont {Chen}(2015)}]{wang15}%
	\BibitemOpen
	\bibfield  {author} {\bibinfo {author} {\bibfnamefont {R.}~\bibnamefont
			{Wang}}\ and\ \bibinfo {author} {\bibfnamefont {L.-W.}\ \bibnamefont
			{Chen}},\ }\href@noop {} {\bibfield  {journal} {\bibinfo  {journal} {Phys.
				Rev. C}\ }\textbf {\bibinfo {volume} {92}},\ \bibinfo {pages} {031303}
		(\bibinfo {year} {2015})}\BibitemShut {NoStop}%
	\bibitem [{\citenamefont {Dong}\ \emph {et~al.}(2011)\citenamefont {Dong},
		\citenamefont {Zuo},\ and\ \citenamefont {Scheid}}]{dong11}%
	\BibitemOpen
	\bibfield  {author} {\bibinfo {author} {\bibfnamefont {J.}~\bibnamefont
			{Dong}}, \bibinfo {author} {\bibfnamefont {W.}~\bibnamefont {Zuo}}, \ and\
		\bibinfo {author} {\bibfnamefont {W.}~\bibnamefont {Scheid}},\ }\href@noop {}
	{\bibfield  {journal} {\bibinfo  {journal} {Phys. Rev. Lett.}\ }\textbf
		{\bibinfo {volume} {107}},\ \bibinfo {pages} {012501} (\bibinfo {year}
		{2011})}\BibitemShut {NoStop}%
	\bibitem [{\citenamefont {Shin}\ \emph {et~al.}(2016)\citenamefont {Shin},
		\citenamefont {Lim}, \citenamefont {Hyun},\ and\ \citenamefont
		{Oh}}]{shin16}%
	\BibitemOpen
	\bibfield  {author} {\bibinfo {author} {\bibfnamefont {E.}~\bibnamefont
			{Shin}}, \bibinfo {author} {\bibfnamefont {Y.}~\bibnamefont {Lim}}, \bibinfo
		{author} {\bibfnamefont {C.~H.}\ \bibnamefont {Hyun}}, \ and\ \bibinfo
		{author} {\bibfnamefont {Y.}~\bibnamefont {Oh}},\ }\href@noop {} {\bibfield
		{journal} {\bibinfo  {journal} {Phys. Rev. C}\ }\textbf {\bibinfo {volume}
			{94}},\ \bibinfo {pages} {024320} (\bibinfo {year} {2016})}\BibitemShut
	{NoStop}%
	\bibitem [{\citenamefont {Lim}\ and\ \citenamefont {Oh}(2017)}]{lim17alfa}%
	\BibitemOpen
	\bibfield  {author} {\bibinfo {author} {\bibfnamefont {Y.}~\bibnamefont
			{Lim}}\ and\ \bibinfo {author} {\bibfnamefont {Y.}~\bibnamefont {Oh}},\
	}\href@noop {} {\bibfield  {journal} {\bibinfo  {journal} {Phys. Rev. C}\
		}\textbf {\bibinfo {volume} {95}},\ \bibinfo {pages} {034311} (\bibinfo
		{year} {2017})}\BibitemShut {NoStop}%
	\bibitem [{\citenamefont {Li}(2002{\natexlab{a}})}]{Li2002a}%
	\BibitemOpen
	\bibfield  {author} {\bibinfo {author} {\bibfnamefont {B.-A.}\ \bibnamefont
			{Li}},\ }\href@noop {} {\bibfield  {journal} {\bibinfo  {journal} {Phys. Rev.
				Lett.}\ }\textbf {\bibinfo {volume} {88}},\ \bibinfo {pages} {192701}
		(\bibinfo {year} {2002}{\natexlab{a}})}\BibitemShut {NoStop}%
	\bibitem [{\citenamefont {Li}(2002{\natexlab{b}})}]{Li2002b}%
	\BibitemOpen
	\bibfield  {author} {\bibinfo {author} {\bibfnamefont {B.-A.}\ \bibnamefont
			{Li}},\ }\href@noop {} {\bibfield  {journal} {\bibinfo  {journal} {Nucl.
				Phys. A}\ }\textbf {\bibinfo {volume} {708}},\ \bibinfo {pages} {365}
		(\bibinfo {year} {2002}{\natexlab{b}})}\BibitemShut {NoStop}%
	\bibitem [{\citenamefont {Tsang}\ \emph {et~al.}(2004)\citenamefont {Tsang},
		\citenamefont {Liu}, \citenamefont {Shi}, \citenamefont {Danielewicz},
		\citenamefont {Gelbke}, \citenamefont {Liu}, \citenamefont {Lynch},
		\citenamefont {Tan}, \citenamefont {Verde}, \citenamefont {Wagner} \emph
		{et~al.}}]{Tsang2004}%
	\BibitemOpen
	\bibfield  {author} {\bibinfo {author} {\bibfnamefont {M.}~\bibnamefont
			{Tsang}}, \bibinfo {author} {\bibfnamefont {T.}~\bibnamefont {Liu}}, \bibinfo
		{author} {\bibfnamefont {L.}~\bibnamefont {Shi}}, \bibinfo {author}
		{\bibfnamefont {P.}~\bibnamefont {Danielewicz}}, \bibinfo {author}
		{\bibfnamefont {C.}~\bibnamefont {Gelbke}}, \bibinfo {author} {\bibfnamefont
			{X.}~\bibnamefont {Liu}}, \bibinfo {author} {\bibfnamefont {W.}~\bibnamefont
			{Lynch}}, \bibinfo {author} {\bibfnamefont {W.}~\bibnamefont {Tan}}, \bibinfo
		{author} {\bibfnamefont {G.}~\bibnamefont {Verde}}, \bibinfo {author}
		{\bibfnamefont {A.}~\bibnamefont {Wagner}},  \emph {et~al.},\ }\href@noop {}
	{\bibfield  {journal} {\bibinfo  {journal} {Phys. Rev. Lett.}\ }\textbf
		{\bibinfo {volume} {92}},\ \bibinfo {pages} {062701} (\bibinfo {year}
		{2004})}\BibitemShut {NoStop}%
	\bibitem [{\citenamefont {Di~Toro}\ \emph {et~al.}(2010)\citenamefont
		{Di~Toro}, \citenamefont {Baran}, \citenamefont {Colonna},\ and\
		\citenamefont {Greco}}]{Di2010}%
	\BibitemOpen
	\bibfield  {author} {\bibinfo {author} {\bibfnamefont {M.}~\bibnamefont
			{Di~Toro}}, \bibinfo {author} {\bibfnamefont {V.}~\bibnamefont {Baran}},
		\bibinfo {author} {\bibfnamefont {M.}~\bibnamefont {Colonna}}, \ and\
		\bibinfo {author} {\bibfnamefont {V.}~\bibnamefont {Greco}},\ }\href@noop {}
	{\bibfield  {journal} {\bibinfo  {journal} {J. Phys. G}\ }\textbf {\bibinfo
			{volume} {37}},\ \bibinfo {pages} {083101} (\bibinfo {year}
		{2010})}\BibitemShut {NoStop}%
	\bibitem [{\citenamefont {Page}\ and\ \citenamefont
		{Applegate}(1992)}]{page1992}%
	\BibitemOpen
	\bibfield  {author} {\bibinfo {author} {\bibfnamefont {D.}~\bibnamefont
			{Page}}\ and\ \bibinfo {author} {\bibfnamefont {J.~H.}\ \bibnamefont
			{Applegate}},\ }\href@noop {} {\bibfield  {journal} {\bibinfo  {journal}
			{Astrophys. J.}\ }\textbf {\bibinfo {volume} {394}},\ \bibinfo {pages} {L17}
		(\bibinfo {year} {1992})}\BibitemShut {NoStop}%
	\bibitem [{\citenamefont {Lim}\ \emph {et~al.}(2017)\citenamefont {Lim},
		\citenamefont {Hyun},\ and\ \citenamefont {Lee}}]{Lim2017a}%
	\BibitemOpen
	\bibfield  {author} {\bibinfo {author} {\bibfnamefont {Y.}~\bibnamefont
			{Lim}}, \bibinfo {author} {\bibfnamefont {C.~H.}\ \bibnamefont {Hyun}}, \
		and\ \bibinfo {author} {\bibfnamefont {C.-H.}\ \bibnamefont {Lee}},\
	}\href@noop {} {\bibfield  {journal} {\bibinfo  {journal} {Int. J. Mod.
				Phys.}\ }\textbf {\bibinfo {volume} {E26}},\ \bibinfo {pages} {1750015}
		(\bibinfo {year} {2017})}\BibitemShut {NoStop}%
	\bibitem [{\citenamefont {Lattimer}\ and\ \citenamefont
		{Prakash}(2001)}]{lattimer01}%
	\BibitemOpen
	\bibfield  {author} {\bibinfo {author} {\bibfnamefont {J.~M.}\ \bibnamefont
			{Lattimer}}\ and\ \bibinfo {author} {\bibfnamefont {M.}~\bibnamefont
			{Prakash}},\ }\href@noop {} {\bibfield  {journal} {\bibinfo  {journal}
			{Astrophys. J.}\ }\textbf {\bibinfo {volume} {550}},\ \bibinfo {pages} {426}
		(\bibinfo {year} {2001})}\BibitemShut {NoStop}%
	\bibitem [{\citenamefont {Gandolfi}\ \emph {et~al.}(2012)\citenamefont
		{Gandolfi}, \citenamefont {Carlson},\ and\ \citenamefont
		{Reddy}}]{gandolfi2012}%
	\BibitemOpen
	\bibfield  {author} {\bibinfo {author} {\bibfnamefont {S.}~\bibnamefont
			{Gandolfi}}, \bibinfo {author} {\bibfnamefont {J.}~\bibnamefont {Carlson}}, \
		and\ \bibinfo {author} {\bibfnamefont {S.}~\bibnamefont {Reddy}},\
	}\href@noop {} {\bibfield  {journal} {\bibinfo  {journal} {Phys. Rev. C}\
		}\textbf {\bibinfo {volume} {85}},\ \bibinfo {pages} {032801} (\bibinfo
		{year} {2012})}\BibitemShut {NoStop}%
	\bibitem [{\citenamefont {Steiner}\ and\ \citenamefont
		{Gandolfi}(2012)}]{steiner2012a}%
	\BibitemOpen
	\bibfield  {author} {\bibinfo {author} {\bibfnamefont {A.}~\bibnamefont
			{Steiner}}\ and\ \bibinfo {author} {\bibfnamefont {S.}~\bibnamefont
			{Gandolfi}},\ }\href@noop {} {\bibfield  {journal} {\bibinfo  {journal}
			{Phys. Rev. Lett.}\ }\textbf {\bibinfo {volume} {108}},\ \bibinfo {pages}
		{081102} (\bibinfo {year} {2012})}\BibitemShut {NoStop}%
	\bibitem [{\citenamefont {Lim}\ \emph {et~al.}(2019)\citenamefont {Lim},
		\citenamefont {Holt},\ and\ \citenamefont {Stahulak}}]{lim2019moi}%
	\BibitemOpen
	\bibfield  {author} {\bibinfo {author} {\bibfnamefont {Y.}~\bibnamefont
			{Lim}}, \bibinfo {author} {\bibfnamefont {J.~W.}\ \bibnamefont {Holt}}, \
		and\ \bibinfo {author} {\bibfnamefont {R.~J.}\ \bibnamefont {Stahulak}},\
	}\href@noop {} {\bibfield  {journal} {\bibinfo  {journal} {Phys. Rev. C}\
		}\textbf {\bibinfo {volume} {100}},\ \bibinfo {pages} {035802} (\bibinfo
		{year} {2019})}\BibitemShut {NoStop}%
	\bibitem [{\citenamefont {Lim}\ and\ \citenamefont {Holt}(2019)}]{lim19e}%
	\BibitemOpen
	\bibfield  {author} {\bibinfo {author} {\bibfnamefont {Y.}~\bibnamefont
			{Lim}}\ and\ \bibinfo {author} {\bibfnamefont {J.~W.}\ \bibnamefont {Holt}},\
	}\href@noop {} {\bibfield  {journal} {\bibinfo  {journal} {Eur. Phys. J. A}\
		}\textbf {\bibinfo {volume} {55}},\ \bibinfo {pages} {209} (\bibinfo {year}
		{2019})}\BibitemShut {NoStop}%
	\bibitem [{\citenamefont {Baldo}\ and\ \citenamefont
		{Burgio}(2016)}]{baldo2016a}%
	\BibitemOpen
	\bibfield  {author} {\bibinfo {author} {\bibfnamefont {M.}~\bibnamefont
			{Baldo}}\ and\ \bibinfo {author} {\bibfnamefont {G.}~\bibnamefont {Burgio}},\
	}\href@noop {} {\bibfield  {journal} {\bibinfo  {journal} {Progress in
				Particle and Nuclear Physics}\ }\textbf {\bibinfo {volume} {91}},\ \bibinfo
		{pages} {203} (\bibinfo {year} {2016})}\BibitemShut {NoStop}%
	\bibitem [{\citenamefont {Newton}\ and\ \citenamefont
		{Crocombe}(2021)}]{newton2021a}%
	\BibitemOpen
	\bibfield  {author} {\bibinfo {author} {\bibfnamefont {W.~G.}\ \bibnamefont
			{Newton}}\ and\ \bibinfo {author} {\bibfnamefont {G.}~\bibnamefont
			{Crocombe}},\ }\href@noop {} {\bibfield  {journal} {\bibinfo  {journal}
			{Phys. Rev. C}\ }\textbf {\bibinfo {volume} {103}},\ \bibinfo {pages}
		{064323} (\bibinfo {year} {2021})}\BibitemShut {NoStop}%
	\bibitem [{\citenamefont {Lattimer}(2023)}]{lattimer2023a}%
	\BibitemOpen
	\bibfield  {author} {\bibinfo {author} {\bibfnamefont {J.~M.}\ \bibnamefont
			{Lattimer}},\ }\href@noop {} {\bibfield  {journal} {\bibinfo  {journal}
			{Particles}\ }\textbf {\bibinfo {volume} {6}},\ \bibinfo {pages} {30}
		(\bibinfo {year} {2023})}\BibitemShut {NoStop}%
	\bibitem [{\citenamefont {Hebeler}\ \emph {et~al.}(2011)\citenamefont
		{Hebeler}, \citenamefont {Bogner}, \citenamefont {Furnstahl}, \citenamefont
		{Nogga},\ and\ \citenamefont {Schwenk}}]{hebeler11}%
	\BibitemOpen
	\bibfield  {author} {\bibinfo {author} {\bibfnamefont {K.}~\bibnamefont
			{Hebeler}}, \bibinfo {author} {\bibfnamefont {S.~K.}\ \bibnamefont {Bogner}},
		\bibinfo {author} {\bibfnamefont {R.~J.}\ \bibnamefont {Furnstahl}}, \bibinfo
		{author} {\bibfnamefont {A.}~\bibnamefont {Nogga}}, \ and\ \bibinfo {author}
		{\bibfnamefont {A.}~\bibnamefont {Schwenk}},\ }\href@noop {} {\bibfield
		{journal} {\bibinfo  {journal} {Phys. Rev. C}\ }\textbf {\bibinfo {volume}
			{83}},\ \bibinfo {pages} {031301} (\bibinfo {year} {2011})}\BibitemShut
	{NoStop}%
	\bibitem [{\citenamefont {Gezerlis}\ \emph {et~al.}(2013)\citenamefont
		{Gezerlis}, \citenamefont {Tews}, \citenamefont {Epelbaum}, \citenamefont
		{Gandolfi}, \citenamefont {Hebeler}, \citenamefont {Nogga},\ and\
		\citenamefont {Schwenk}}]{gezerlis13}%
	\BibitemOpen
	\bibfield  {author} {\bibinfo {author} {\bibfnamefont {A.}~\bibnamefont
			{Gezerlis}}, \bibinfo {author} {\bibfnamefont {I.}~\bibnamefont {Tews}},
		\bibinfo {author} {\bibfnamefont {E.}~\bibnamefont {Epelbaum}}, \bibinfo
		{author} {\bibfnamefont {S.}~\bibnamefont {Gandolfi}}, \bibinfo {author}
		{\bibfnamefont {K.}~\bibnamefont {Hebeler}}, \bibinfo {author} {\bibfnamefont
			{A.}~\bibnamefont {Nogga}}, \ and\ \bibinfo {author} {\bibfnamefont
			{A.}~\bibnamefont {Schwenk}},\ }\href@noop {} {\bibfield  {journal} {\bibinfo
			{journal} {Phys. Rev. Lett.}\ }\textbf {\bibinfo {volume} {111}},\ \bibinfo
		{pages} {032501} (\bibinfo {year} {2013})}\BibitemShut {NoStop}%
	\bibitem [{\citenamefont {Roggero}\ \emph {et~al.}(2014)\citenamefont
		{Roggero}, \citenamefont {Mukherjee},\ and\ \citenamefont
		{Pederiva}}]{roggero14}%
	\BibitemOpen
	\bibfield  {author} {\bibinfo {author} {\bibfnamefont {A.}~\bibnamefont
			{Roggero}}, \bibinfo {author} {\bibfnamefont {A.}~\bibnamefont {Mukherjee}},
		\ and\ \bibinfo {author} {\bibfnamefont {F.}~\bibnamefont {Pederiva}},\
	}\href@noop {} {\bibfield  {journal} {\bibinfo  {journal} {Phys. Rev. Lett.}\
		}\textbf {\bibinfo {volume} {112}},\ \bibinfo {pages} {221103} (\bibinfo
		{year} {2014})}\BibitemShut {NoStop}%
	\bibitem [{\citenamefont {Wlaz{\l}owski}\ \emph {et~al.}(2014)\citenamefont
		{Wlaz{\l}owski}, \citenamefont {Holt}, \citenamefont {Moroz}, \citenamefont
		{Bulgac},\ and\ \citenamefont {Roche}}]{wlazlowski14}%
	\BibitemOpen
	\bibfield  {author} {\bibinfo {author} {\bibfnamefont {G.}~\bibnamefont
			{Wlaz{\l}owski}}, \bibinfo {author} {\bibfnamefont {J.}~\bibnamefont {Holt}},
		\bibinfo {author} {\bibfnamefont {S.}~\bibnamefont {Moroz}}, \bibinfo
		{author} {\bibfnamefont {A.}~\bibnamefont {Bulgac}}, \ and\ \bibinfo {author}
		{\bibfnamefont {K.}~\bibnamefont {Roche}},\ }\href@noop {} {\bibfield
		{journal} {\bibinfo  {journal} {Phys. Rev. Lett.}\ }\textbf {\bibinfo
			{volume} {113}},\ \bibinfo {pages} {182503} (\bibinfo {year}
		{2014})}\BibitemShut {NoStop}%
	\bibitem [{\citenamefont {Drischler}\ \emph {et~al.}(2014)\citenamefont
		{Drischler}, \citenamefont {Som\`a},\ and\ \citenamefont
		{Schwenk}}]{drischler14}%
	\BibitemOpen
	\bibfield  {author} {\bibinfo {author} {\bibfnamefont {C.}~\bibnamefont
			{Drischler}}, \bibinfo {author} {\bibfnamefont {V.}~\bibnamefont {Som\`a}}, \
		and\ \bibinfo {author} {\bibfnamefont {A.}~\bibnamefont {Schwenk}},\
	}\href@noop {} {\bibfield  {journal} {\bibinfo  {journal} {Phys. Rev. C}\
		}\textbf {\bibinfo {volume} {89}},\ \bibinfo {pages} {025806} (\bibinfo
		{year} {2014})}\BibitemShut {NoStop}%
	\bibitem [{\citenamefont {Drischler}\ \emph
		{et~al.}(2016{\natexlab{a}})\citenamefont {Drischler}, \citenamefont
		{Hebeler},\ and\ \citenamefont {Schwenk}}]{drischler15}%
	\BibitemOpen
	\bibfield  {author} {\bibinfo {author} {\bibfnamefont {C.}~\bibnamefont
			{Drischler}}, \bibinfo {author} {\bibfnamefont {K.}~\bibnamefont {Hebeler}},
		\ and\ \bibinfo {author} {\bibfnamefont {A.}~\bibnamefont {Schwenk}},\
	}\href@noop {} {\bibfield  {journal} {\bibinfo  {journal} {Phys. Rev.}\
		}\textbf {\bibinfo {volume} {C93}},\ \bibinfo {pages} {054314} (\bibinfo
		{year} {2016}{\natexlab{a}})}\BibitemShut {NoStop}%
	\bibitem [{\citenamefont {Drischler}\ \emph
		{et~al.}(2016{\natexlab{b}})\citenamefont {Drischler}, \citenamefont
		{Carbone}, \citenamefont {Hebeler},\ and\ \citenamefont
		{Schwenk}}]{drischler16}%
	\BibitemOpen
	\bibfield  {author} {\bibinfo {author} {\bibfnamefont {C.}~\bibnamefont
			{Drischler}}, \bibinfo {author} {\bibfnamefont {A.}~\bibnamefont {Carbone}},
		\bibinfo {author} {\bibfnamefont {K.}~\bibnamefont {Hebeler}}, \ and\
		\bibinfo {author} {\bibfnamefont {A.}~\bibnamefont {Schwenk}},\ }\href@noop
	{} {\bibfield  {journal} {\bibinfo  {journal} {Phys. Rev. C}\ }\textbf
		{\bibinfo {volume} {94}},\ \bibinfo {pages} {054307} (\bibinfo {year}
		{2016}{\natexlab{b}})}\BibitemShut {NoStop}%
	\bibitem [{\citenamefont {Tews}\ \emph {et~al.}(2016)\citenamefont {Tews},
		\citenamefont {Gandolfi}, \citenamefont {Gezerlis},\ and\ \citenamefont
		{Schwenk}}]{tews16}%
	\BibitemOpen
	\bibfield  {author} {\bibinfo {author} {\bibfnamefont {I.}~\bibnamefont
			{Tews}}, \bibinfo {author} {\bibfnamefont {S.}~\bibnamefont {Gandolfi}},
		\bibinfo {author} {\bibfnamefont {A.}~\bibnamefont {Gezerlis}}, \ and\
		\bibinfo {author} {\bibfnamefont {A.}~\bibnamefont {Schwenk}},\ }\href@noop
	{} {\bibfield  {journal} {\bibinfo  {journal} {Phys. Rev. C}\ }\textbf
		{\bibinfo {volume} {93}},\ \bibinfo {pages} {024305} (\bibinfo {year}
		{2016})}\BibitemShut {NoStop}%
	\bibitem [{\citenamefont {Wellenhofer}\ \emph {et~al.}(2016)\citenamefont
		{Wellenhofer}, \citenamefont {Holt},\ and\ \citenamefont
		{Kaiser}}]{Holt:2016pjb}%
	\BibitemOpen
	\bibfield  {author} {\bibinfo {author} {\bibfnamefont {C.}~\bibnamefont
			{Wellenhofer}}, \bibinfo {author} {\bibfnamefont {J.~W.}\ \bibnamefont
			{Holt}}, \ and\ \bibinfo {author} {\bibfnamefont {N.}~\bibnamefont
			{Kaiser}},\ }\href@noop {} {\bibfield  {journal} {\bibinfo  {journal} {Phys.
				Rev. C}\ }\textbf {\bibinfo {volume} {93}},\ \bibinfo {pages} {055802}
		(\bibinfo {year} {2016})}\BibitemShut {NoStop}%
	\bibitem [{\citenamefont {Holt}\ \emph {et~al.}(2012)\citenamefont {Holt},
		\citenamefont {Kaiser},\ and\ \citenamefont {Weise}}]{Holt:2011yj}%
	\BibitemOpen
	\bibfield  {author} {\bibinfo {author} {\bibfnamefont {J.~W.}\ \bibnamefont
			{Holt}}, \bibinfo {author} {\bibfnamefont {N.}~\bibnamefont {Kaiser}}, \ and\
		\bibinfo {author} {\bibfnamefont {W.}~\bibnamefont {Weise}},\ }\href@noop {}
	{\bibfield  {journal} {\bibinfo  {journal} {Nucl. Phys. A}\ }\textbf
		{\bibinfo {volume} {876}},\ \bibinfo {pages} {61} (\bibinfo {year}
		{2012})}\BibitemShut {NoStop}%
	\bibitem [{\citenamefont {Holt}\ and\ \citenamefont {Lim}(2018)}]{holt2018b}%
	\BibitemOpen
	\bibfield  {author} {\bibinfo {author} {\bibfnamefont {J.~W.}\ \bibnamefont
			{Holt}}\ and\ \bibinfo {author} {\bibfnamefont {Y.}~\bibnamefont {Lim}},\
	}\href@noop {} {\bibfield  {journal} {\bibinfo  {journal} {Phys. Lett.}\
		}\textbf {\bibinfo {volume} {B784}},\ \bibinfo {pages} {77} (\bibinfo {year}
		{2018})}\BibitemShut {NoStop}%
	\bibitem [{\citenamefont {Abbott}\ \emph
		{et~al.}(2017{\natexlab{a}})\citenamefont {Abbott} \emph
		{et~al.}}]{gw170817}%
	\BibitemOpen
	\bibfield  {author} {\bibinfo {author} {\bibfnamefont {B.~P.}\ \bibnamefont
			{Abbott}} \emph {et~al.} (\bibinfo {collaboration} {LIGO Scientific
			Collaboration and Virgo Collaboration}),\ }\href@noop {} {\bibfield
		{journal} {\bibinfo  {journal} {Phys. Rev. Lett.}\ }\textbf {\bibinfo
			{volume} {119}},\ \bibinfo {pages} {161101} (\bibinfo {year}
		{2017}{\natexlab{a}})}\BibitemShut {NoStop}%
	\bibitem [{\citenamefont {Abbott}\ \emph
		{et~al.}(2017{\natexlab{b}})\citenamefont {Abbott} \emph
		{et~al.}}]{gw170817apj}%
	\BibitemOpen
	\bibfield  {author} {\bibinfo {author} {\bibfnamefont {B.~P.}\ \bibnamefont
			{Abbott}} \emph {et~al.},\ }\href@noop {} {\bibfield  {journal} {\bibinfo
			{journal} {The Astrophysical Journal Letters}\ }\textbf {\bibinfo {volume}
			{848}},\ \bibinfo {pages} {L12} (\bibinfo {year}
		{2017}{\natexlab{b}})}\BibitemShut {NoStop}%
	\bibitem [{\citenamefont {De}\ \emph {et~al.}(2018)\citenamefont {De},
		\citenamefont {Finstad}, \citenamefont {Lattimer}, \citenamefont {Brown},
		\citenamefont {Berger},\ and\ \citenamefont {Biwer}}]{de18}%
	\BibitemOpen
	\bibfield  {author} {\bibinfo {author} {\bibfnamefont {S.}~\bibnamefont
			{De}}, \bibinfo {author} {\bibfnamefont {D.}~\bibnamefont {Finstad}},
		\bibinfo {author} {\bibfnamefont {J.~M.}\ \bibnamefont {Lattimer}}, \bibinfo
		{author} {\bibfnamefont {D.~A.}\ \bibnamefont {Brown}}, \bibinfo {author}
		{\bibfnamefont {E.}~\bibnamefont {Berger}}, \ and\ \bibinfo {author}
		{\bibfnamefont {C.~M.}\ \bibnamefont {Biwer}},\ }\href@noop {} {\bibfield
		{journal} {\bibinfo  {journal} {Phys. Rev. Lett.}\ }\textbf {\bibinfo
			{volume} {121}},\ \bibinfo {pages} {091102} (\bibinfo {year}
		{2018})}\BibitemShut {NoStop}%
	\bibitem [{\citenamefont {Abbott}\ \emph {et~al.}(2018)\citenamefont {Abbott}
		\emph {et~al.}}]{ligo18a}%
	\BibitemOpen
	\bibfield  {author} {\bibinfo {author} {\bibfnamefont {B.~P.}\ \bibnamefont
			{Abbott}} \emph {et~al.} (\bibinfo {collaboration} {The LIGO Scientific
			Collaboration and the Virgo Collaboration}),\ }\href@noop {} {\bibfield
		{journal} {\bibinfo  {journal} {Phys. Rev. Lett.}\ }\textbf {\bibinfo
			{volume} {121}},\ \bibinfo {pages} {161101} (\bibinfo {year}
		{2018})}\BibitemShut {NoStop}%
	\bibitem [{\citenamefont {Fattoyev}\ \emph {et~al.}(2018)\citenamefont
		{Fattoyev}, \citenamefont {Piekarewicz},\ and\ \citenamefont
		{Horowitz}}]{fattoyev18}%
	\BibitemOpen
	\bibfield  {author} {\bibinfo {author} {\bibfnamefont {F.~J.}\ \bibnamefont
			{Fattoyev}}, \bibinfo {author} {\bibfnamefont {J.}~\bibnamefont
			{Piekarewicz}}, \ and\ \bibinfo {author} {\bibfnamefont {C.~J.}\ \bibnamefont
			{Horowitz}},\ }\href@noop {} {\bibfield  {journal} {\bibinfo  {journal}
			{Phys. Rev. Lett.}\ }\textbf {\bibinfo {volume} {120}},\ \bibinfo {pages}
		{172702} (\bibinfo {year} {2018})}\BibitemShut {NoStop}%
	\bibitem [{\citenamefont {Krastev}\ and\ \citenamefont
		{Li}(2019)}]{Krastev2018}%
	\BibitemOpen
	\bibfield  {author} {\bibinfo {author} {\bibfnamefont {P.~G.}\ \bibnamefont
			{Krastev}}\ and\ \bibinfo {author} {\bibfnamefont {B.-A.}\ \bibnamefont
			{Li}},\ }\href@noop {} {\bibfield  {journal} {\bibinfo  {journal} {J. Phys.
				G}\ }\textbf {\bibinfo {volume} {46}},\ \bibinfo {pages} {074001} (\bibinfo
		{year} {2019})}\BibitemShut {NoStop}%
	\bibitem [{\citenamefont {Lim}\ and\ \citenamefont {Holt}(2018)}]{Lim2018}%
	\BibitemOpen
	\bibfield  {author} {\bibinfo {author} {\bibfnamefont {Y.}~\bibnamefont
			{Lim}}\ and\ \bibinfo {author} {\bibfnamefont {J.~W.}\ \bibnamefont {Holt}},\
	}\href@noop {} {\bibfield  {journal} {\bibinfo  {journal} {Phys. Rev. Lett.}\
		}\textbf {\bibinfo {volume} {121}},\ \bibinfo {pages} {062701} (\bibinfo
		{year} {2018})}\BibitemShut {NoStop}%
	\bibitem [{\citenamefont {Zhang}\ and\ \citenamefont
		{Li}(2019{\natexlab{a}})}]{Zhang2018a}%
	\BibitemOpen
	\bibfield  {author} {\bibinfo {author} {\bibfnamefont {N.-B.}\ \bibnamefont
			{Zhang}}\ and\ \bibinfo {author} {\bibfnamefont {B.-A.}\ \bibnamefont {Li}},\
	}\href@noop {} {\bibfield  {journal} {\bibinfo  {journal} {Eur. Phys. J. A}\
		}\textbf {\bibinfo {volume} {55}},\ \bibinfo {pages} {39} (\bibinfo {year}
		{2019}{\natexlab{a}})}\BibitemShut {NoStop}%
	\bibitem [{\citenamefont {Zhang}\ and\ \citenamefont
		{Li}(2019{\natexlab{b}})}]{Zhang2018b}%
	\BibitemOpen
	\bibfield  {author} {\bibinfo {author} {\bibfnamefont {N.-B.}\ \bibnamefont
			{Zhang}}\ and\ \bibinfo {author} {\bibfnamefont {B.-A.}\ \bibnamefont {Li}},\
	}\href@noop {} {\bibfield  {journal} {\bibinfo  {journal} {J. Phys. G}\
		}\textbf {\bibinfo {volume} {46}},\ \bibinfo {pages} {014002} (\bibinfo
		{year} {2019}{\natexlab{b}})}\BibitemShut {NoStop}%
	\bibitem [{\citenamefont {Myers}\ and\ \citenamefont
		{Swiatecki}(1966)}]{Myers1966}%
	\BibitemOpen
	\bibfield  {author} {\bibinfo {author} {\bibfnamefont {W.}~\bibnamefont
			{Myers}}\ and\ \bibinfo {author} {\bibfnamefont {W.}~\bibnamefont
			{Swiatecki}},\ }\href@noop {} {\bibfield  {journal} {\bibinfo  {journal}
			{Nuclear Physics}\ }\textbf {\bibinfo {volume} {81}},\ \bibinfo {pages} {1}
		(\bibinfo {year} {1966})}\BibitemShut {NoStop}%
	\bibitem [{\citenamefont {Duflo}\ and\ \citenamefont {Zuker}(1995)}]{DZ1995}%
	\BibitemOpen
	\bibfield  {author} {\bibinfo {author} {\bibfnamefont {J.}~\bibnamefont
			{Duflo}}\ and\ \bibinfo {author} {\bibfnamefont {A.}~\bibnamefont {Zuker}},\
	}\href@noop {} {\bibfield  {journal} {\bibinfo  {journal} {Phys. Rev. C}\
		}\textbf {\bibinfo {volume} {52}},\ \bibinfo {pages} {R23} (\bibinfo {year}
		{1995})}\BibitemShut {NoStop}%
	\bibitem [{\citenamefont {M\"oller}\ \emph {et~al.}(2012)\citenamefont
		{M\"oller}, \citenamefont {Myers}, \citenamefont {Sagawa},\ and\
		\citenamefont {Yoshida}}]{frdm12}%
	\BibitemOpen
	\bibfield  {author} {\bibinfo {author} {\bibfnamefont {P.}~\bibnamefont
			{M\"oller}}, \bibinfo {author} {\bibfnamefont {W.~D.}\ \bibnamefont {Myers}},
		\bibinfo {author} {\bibfnamefont {H.}~\bibnamefont {Sagawa}}, \ and\ \bibinfo
		{author} {\bibfnamefont {S.}~\bibnamefont {Yoshida}},\ }\href@noop {}
	{\bibfield  {journal} {\bibinfo  {journal} {Phys. Rev. Lett.}\ }\textbf
		{\bibinfo {volume} {108}},\ \bibinfo {pages} {052501} (\bibinfo {year}
		{2012})}\BibitemShut {NoStop}%
	\bibitem [{\citenamefont {Wang}\ \emph {et~al.}(2014)\citenamefont {Wang},
		\citenamefont {Liu}, \citenamefont {Wu},\ and\ \citenamefont
		{Meng}}]{Wang2014}%
	\BibitemOpen
	\bibfield  {author} {\bibinfo {author} {\bibfnamefont {N.}~\bibnamefont
			{Wang}}, \bibinfo {author} {\bibfnamefont {M.}~\bibnamefont {Liu}}, \bibinfo
		{author} {\bibfnamefont {X.}~\bibnamefont {Wu}}, \ and\ \bibinfo {author}
		{\bibfnamefont {J.}~\bibnamefont {Meng}},\ }\href@noop {} {\bibfield
		{journal} {\bibinfo  {journal} {Phys. Lett. B}\ }\textbf {\bibinfo {volume}
			{734}},\ \bibinfo {pages} {215} (\bibinfo {year} {2014})}\BibitemShut
	{NoStop}%
	\bibitem [{\citenamefont {Steiner}\ \emph {et~al.}(2005)\citenamefont
		{Steiner}, \citenamefont {Prakash}, \citenamefont {Lattimer},\ and\
		\citenamefont {Ellis}}]{Steiner2005325}%
	\BibitemOpen
	\bibfield  {author} {\bibinfo {author} {\bibfnamefont {A.}~\bibnamefont
			{Steiner}}, \bibinfo {author} {\bibfnamefont {M.}~\bibnamefont {Prakash}},
		\bibinfo {author} {\bibfnamefont {J.}~\bibnamefont {Lattimer}}, \ and\
		\bibinfo {author} {\bibfnamefont {P.}~\bibnamefont {Ellis}},\ }\href@noop {}
	{\bibfield  {journal} {\bibinfo  {journal} {Physics Reports}\ }\textbf
		{\bibinfo {volume} {411}},\ \bibinfo {pages} {325 } (\bibinfo {year}
		{2005})}\BibitemShut {NoStop}%
	\bibitem [{\citenamefont {{Dieperink, A. E. L.}}\ and\ \citenamefont {{Van
				Isacker, P.}}(2009)}]{DI2009}%
	\BibitemOpen
	\bibfield  {author} {\bibinfo {author} {\bibnamefont {{Dieperink, A. E.
					L.}}}\ and\ \bibinfo {author} {\bibnamefont {{Van Isacker, P.}}},\
	}\href@noop {} {\bibfield  {journal} {\bibinfo  {journal} {Eur. Phys. J. A}\
		}\textbf {\bibinfo {volume} {42}},\ \bibinfo {pages} {269} (\bibinfo {year}
		{2009})}\BibitemShut {NoStop}%
	\bibitem [{\citenamefont {Antonov}\ \emph {et~al.}(2005)\citenamefont
		{Antonov}, \citenamefont {Kadrev}, \citenamefont {Gaidarov}, \citenamefont
		{Guerra}, \citenamefont {Sarriguren}, \citenamefont {Udias}, \citenamefont
		{Lukyanov}, \citenamefont {Zemlyanaya},\ and\ \citenamefont
		{Krumova}}]{antonov05}%
	\BibitemOpen
	\bibfield  {author} {\bibinfo {author} {\bibfnamefont {A.~N.}\ \bibnamefont
			{Antonov}}, \bibinfo {author} {\bibfnamefont {D.~N.}\ \bibnamefont {Kadrev}},
		\bibinfo {author} {\bibfnamefont {M.~K.}\ \bibnamefont {Gaidarov}}, \bibinfo
		{author} {\bibfnamefont {E.~M.~d.}\ \bibnamefont {Guerra}}, \bibinfo {author}
		{\bibfnamefont {P.}~\bibnamefont {Sarriguren}}, \bibinfo {author}
		{\bibfnamefont {J.~M.}\ \bibnamefont {Udias}}, \bibinfo {author}
		{\bibfnamefont {V.~K.}\ \bibnamefont {Lukyanov}}, \bibinfo {author}
		{\bibfnamefont {E.~V.}\ \bibnamefont {Zemlyanaya}}, \ and\ \bibinfo {author}
		{\bibfnamefont {G.~Z.}\ \bibnamefont {Krumova}},\ }\href@noop {} {\bibfield
		{journal} {\bibinfo  {journal} {Phys. Rev. C}\ }\textbf {\bibinfo {volume}
			{72}},\ \bibinfo {pages} {044307} (\bibinfo {year} {2005})}\BibitemShut
	{NoStop}%
	\bibitem [{\citenamefont {Hatakeyama}\ \emph {et~al.}(2018)\citenamefont
		{Hatakeyama}, \citenamefont {Horiuchi},\ and\ \citenamefont
		{Kohama}}]{hatakeyama18}%
	\BibitemOpen
	\bibfield  {author} {\bibinfo {author} {\bibfnamefont {S.}~\bibnamefont
			{Hatakeyama}}, \bibinfo {author} {\bibfnamefont {W.}~\bibnamefont
			{Horiuchi}}, \ and\ \bibinfo {author} {\bibfnamefont {A.}~\bibnamefont
			{Kohama}},\ }\href@noop {} {\bibfield  {journal} {\bibinfo  {journal} {Phys.
				Rev. C}\ }\textbf {\bibinfo {volume} {97}},\ \bibinfo {pages} {054607}
		(\bibinfo {year} {2018})}\BibitemShut {NoStop}%
	\bibitem [{\citenamefont {Choudhary}\ \emph {et~al.}(2021)\citenamefont
		{Choudhary}, \citenamefont {Horiuchi}, \citenamefont {Kimura},\ and\
		\citenamefont {Chatterjee}}]{choudhary21}%
	\BibitemOpen
	\bibfield  {author} {\bibinfo {author} {\bibfnamefont {V.}~\bibnamefont
			{Choudhary}}, \bibinfo {author} {\bibfnamefont {W.}~\bibnamefont {Horiuchi}},
		\bibinfo {author} {\bibfnamefont {M.}~\bibnamefont {Kimura}}, \ and\ \bibinfo
		{author} {\bibfnamefont {R.}~\bibnamefont {Chatterjee}},\ }\href@noop {}
	{\bibfield  {journal} {\bibinfo  {journal} {Phys. Rev. C}\ }\textbf {\bibinfo
			{volume} {104}},\ \bibinfo {pages} {054313} (\bibinfo {year}
		{2021})}\BibitemShut {NoStop}%
	\bibitem [{\citenamefont {Huang}\ \emph {et~al.}(2017)\citenamefont {Huang},
		\citenamefont {Audi}, \citenamefont {Wang}, \citenamefont {Kondev},
		\citenamefont {Naimi},\ and\ \citenamefont {Xu}}]{ame2016}%
	\BibitemOpen
	\bibfield  {author} {\bibinfo {author} {\bibfnamefont {W.}~\bibnamefont
			{Huang}}, \bibinfo {author} {\bibfnamefont {G.}~\bibnamefont {Audi}},
		\bibinfo {author} {\bibfnamefont {M.}~\bibnamefont {Wang}}, \bibinfo {author}
		{\bibfnamefont {F.~G.}\ \bibnamefont {Kondev}}, \bibinfo {author}
		{\bibfnamefont {S.}~\bibnamefont {Naimi}}, \ and\ \bibinfo {author}
		{\bibfnamefont {X.}~\bibnamefont {Xu}},\ }\href@noop {} {\bibfield  {journal}
		{\bibinfo  {journal} {Chinese Physics C}\ }\textbf {\bibinfo {volume} {41}},\
		\bibinfo {pages} {030002} (\bibinfo {year} {2017})}\BibitemShut {NoStop}%
	\bibitem [{\citenamefont {Goriely}\ \emph {et~al.}(2013)\citenamefont
		{Goriely}, \citenamefont {Chamel},\ and\ \citenamefont {Pearson}}]{hfb24}%
	\BibitemOpen
	\bibfield  {author} {\bibinfo {author} {\bibfnamefont {S.}~\bibnamefont
			{Goriely}}, \bibinfo {author} {\bibfnamefont {N.}~\bibnamefont {Chamel}}, \
		and\ \bibinfo {author} {\bibfnamefont {J.~M.}\ \bibnamefont {Pearson}},\
	}\href@noop {} {\bibfield  {journal} {\bibinfo  {journal} {Phys. Rev. C}\
		}\textbf {\bibinfo {volume} {88}},\ \bibinfo {pages} {024308} (\bibinfo
		{year} {2013})}\BibitemShut {NoStop}%
	\bibitem [{\citenamefont {Geng}\ \emph {et~al.}(2004)\citenamefont {Geng},
		\citenamefont {Toki},\ and\ \citenamefont {Meng}}]{geng04}%
	\BibitemOpen
	\bibfield  {author} {\bibinfo {author} {\bibfnamefont {L.}~\bibnamefont
			{Geng}}, \bibinfo {author} {\bibfnamefont {H.}~\bibnamefont {Toki}}, \ and\
		\bibinfo {author} {\bibfnamefont {J.}~\bibnamefont {Meng}},\ }\href@noop {}
	{\bibfield  {journal} {\bibinfo  {journal} {Progress of Theoretical Physics}\
		}\textbf {\bibinfo {volume} {112}},\ \bibinfo {pages} {603} (\bibinfo {year}
		{2004})}\BibitemShut {NoStop}%
	\bibitem [{\citenamefont {Stoitsov}\ \emph {et~al.}(2005)\citenamefont
		{Stoitsov}, \citenamefont {Dobaczewski}, \citenamefont {Nazarewicz},\ and\
		\citenamefont {Ring}}]{stoitsov05}%
	\BibitemOpen
	\bibfield  {author} {\bibinfo {author} {\bibfnamefont {M.}~\bibnamefont
			{Stoitsov}}, \bibinfo {author} {\bibfnamefont {J.}~\bibnamefont
			{Dobaczewski}}, \bibinfo {author} {\bibfnamefont {W.}~\bibnamefont
			{Nazarewicz}}, \ and\ \bibinfo {author} {\bibfnamefont {P.}~\bibnamefont
			{Ring}},\ }\href@noop {} {\bibfield  {journal} {\bibinfo  {journal} {Computer
				Physics Communications}\ }\textbf {\bibinfo {volume} {167}},\ \bibinfo
		{pages} {43 } (\bibinfo {year} {2005})}\BibitemShut {NoStop}%
	\bibitem [{\citenamefont {Goriely}\ \emph {et~al.}(2009)\citenamefont
		{Goriely}, \citenamefont {Hilaire}, \citenamefont {Girod},\ and\
		\citenamefont {P\'eru}}]{goriely09}%
	\BibitemOpen
	\bibfield  {author} {\bibinfo {author} {\bibfnamefont {S.}~\bibnamefont
			{Goriely}}, \bibinfo {author} {\bibfnamefont {S.}~\bibnamefont {Hilaire}},
		\bibinfo {author} {\bibfnamefont {M.}~\bibnamefont {Girod}}, \ and\ \bibinfo
		{author} {\bibfnamefont {S.}~\bibnamefont {P\'eru}},\ }\href@noop {}
	{\bibfield  {journal} {\bibinfo  {journal} {Phys. Rev. Lett.}\ }\textbf
		{\bibinfo {volume} {102}},\ \bibinfo {pages} {242501} (\bibinfo {year}
		{2009})}\BibitemShut {NoStop}%
	\bibitem [{\citenamefont {Holt}\ \emph
		{et~al.}(2013{\natexlab{a}})\citenamefont {Holt}, \citenamefont
		{Men\'endez},\ and\ \citenamefont {Schwenk}}]{jdholt13}%
	\BibitemOpen
	\bibfield  {author} {\bibinfo {author} {\bibfnamefont {J.~D.}\ \bibnamefont
			{Holt}}, \bibinfo {author} {\bibfnamefont {J.}~\bibnamefont {Men\'endez}}, \
		and\ \bibinfo {author} {\bibfnamefont {A.}~\bibnamefont {Schwenk}},\
	}\href@noop {} {\bibfield  {journal} {\bibinfo  {journal} {Phys. Rev. Lett.}\
		}\textbf {\bibinfo {volume} {110}},\ \bibinfo {pages} {022502} (\bibinfo
		{year} {2013}{\natexlab{a}})}\BibitemShut {NoStop}%
	\bibitem [{\citenamefont {Hagen}\ \emph
		{et~al.}(2014{\natexlab{a}})\citenamefont {Hagen}, \citenamefont
		{Papenbrock}, \citenamefont {Hjorth-Jensen},\ and\ \citenamefont
		{Dean}}]{Hagen14}%
	\BibitemOpen
	\bibfield  {author} {\bibinfo {author} {\bibfnamefont {G.}~\bibnamefont
			{Hagen}}, \bibinfo {author} {\bibfnamefont {T.}~\bibnamefont {Papenbrock}},
		\bibinfo {author} {\bibfnamefont {M.}~\bibnamefont {Hjorth-Jensen}}, \ and\
		\bibinfo {author} {\bibfnamefont {D.~J.}\ \bibnamefont {Dean}},\ }\href@noop
	{} {\bibfield  {journal} {\bibinfo  {journal} {Reports on Progress in
				Physics}\ }\textbf {\bibinfo {volume} {77}},\ \bibinfo {pages} {096302}
		(\bibinfo {year} {2014}{\natexlab{a}})}\BibitemShut {NoStop}%
	\bibitem [{\citenamefont {Hergert}\ \emph {et~al.}(2016)\citenamefont
		{Hergert}, \citenamefont {Bogner}, \citenamefont {Morris}, \citenamefont
		{Schwenk},\ and\ \citenamefont {Tsukiyama}}]{Hergert16}%
	\BibitemOpen
	\bibfield  {author} {\bibinfo {author} {\bibfnamefont {H.}~\bibnamefont
			{Hergert}}, \bibinfo {author} {\bibfnamefont {S.}~\bibnamefont {Bogner}},
		\bibinfo {author} {\bibfnamefont {T.}~\bibnamefont {Morris}}, \bibinfo
		{author} {\bibfnamefont {A.}~\bibnamefont {Schwenk}}, \ and\ \bibinfo
		{author} {\bibfnamefont {K.}~\bibnamefont {Tsukiyama}},\ }\href@noop {}
	{\bibfield  {journal} {\bibinfo  {journal} {Physics Reports}\ }\textbf
		{\bibinfo {volume} {621}},\ \bibinfo {pages} {165 } (\bibinfo {year}
		{2016})}\BibitemShut {NoStop}%
	\bibitem [{\citenamefont {Bennaceur}\ and\ \citenamefont
		{Dobaczewski}(2005)}]{Bennaceur05}%
	\BibitemOpen
	\bibfield  {author} {\bibinfo {author} {\bibfnamefont {K.}~\bibnamefont
			{Bennaceur}}\ and\ \bibinfo {author} {\bibfnamefont {J.}~\bibnamefont
			{Dobaczewski}},\ }\href@noop {} {\bibfield  {journal} {\bibinfo  {journal}
			{Computer Physics Communications}\ }\textbf {\bibinfo {volume} {168}},\
		\bibinfo {pages} {96} (\bibinfo {year} {2005})}\BibitemShut {NoStop}%
	\bibitem [{\citenamefont {Oyamatsu}\ and\ \citenamefont
		{Iida}(2007)}]{Oyamatsu2007}%
	\BibitemOpen
	\bibfield  {author} {\bibinfo {author} {\bibfnamefont {K.}~\bibnamefont
			{Oyamatsu}}\ and\ \bibinfo {author} {\bibfnamefont {K.}~\bibnamefont
			{Iida}},\ }\href@noop {} {\bibfield  {journal} {\bibinfo  {journal} {Phys.
				Rev. C}\ }\textbf {\bibinfo {volume} {75}},\ \bibinfo {pages} {015801}
		(\bibinfo {year} {2007})}\BibitemShut {NoStop}%
	\bibitem [{\citenamefont {Ravenhall}\ \emph {et~al.}(1983)\citenamefont
		{Ravenhall}, \citenamefont {Pethick},\ and\ \citenamefont
		{Lattimer}}]{RPL1983}%
	\BibitemOpen
	\bibfield  {author} {\bibinfo {author} {\bibfnamefont {D.}~\bibnamefont
			{Ravenhall}}, \bibinfo {author} {\bibfnamefont {C.}~\bibnamefont {Pethick}},
		\ and\ \bibinfo {author} {\bibfnamefont {J.}~\bibnamefont {Lattimer}},\
	}\href@noop {} {\bibfield  {journal} {\bibinfo  {journal} {Nucl. Phys. A}\
		}\textbf {\bibinfo {volume} {407}},\ \bibinfo {pages} {571 } (\bibinfo {year}
		{1983})}\BibitemShut {NoStop}%
	\bibitem [{\citenamefont {{Lattimer}}\ and\ \citenamefont
		{{Swesty}}(1991)}]{ls1991}%
	\BibitemOpen
	\bibfield  {author} {\bibinfo {author} {\bibfnamefont {J.~M.}\ \bibnamefont
			{{Lattimer}}}\ and\ \bibinfo {author} {\bibfnamefont {D.~F.}\ \bibnamefont
			{{Swesty}}},\ }\href@noop {} {\bibfield  {journal} {\bibinfo  {journal}
			{Nucl. Phys. A}\ }\textbf {\bibinfo {volume} {535}},\ \bibinfo {pages} {331}
		(\bibinfo {year} {1991})}\BibitemShut {NoStop}%
	\bibitem [{\citenamefont {Lim}\ and\ \citenamefont {Holt}(2017)}]{lim2017c}%
	\BibitemOpen
	\bibfield  {author} {\bibinfo {author} {\bibfnamefont {Y.}~\bibnamefont
			{Lim}}\ and\ \bibinfo {author} {\bibfnamefont {J.~W.}\ \bibnamefont {Holt}},\
	}\href@noop {} {\bibfield  {journal} {\bibinfo  {journal} {Phys. Rev. C}\
		}\textbf {\bibinfo {volume} {95}},\ \bibinfo {pages} {065805} (\bibinfo
		{year} {2017})}\BibitemShut {NoStop}%
	\bibitem [{\citenamefont {Dutra}\ \emph {et~al.}(2012)\citenamefont {Dutra},
		\citenamefont {Louren\ifmmode~\mbox{\c{c}}\else \c{c}\fi{}o}, \citenamefont
		{S\'a~Martins}, \citenamefont {Delfino}, \citenamefont {Stone},\ and\
		\citenamefont {Stevenson}}]{dutra2012}%
	\BibitemOpen
	\bibfield  {author} {\bibinfo {author} {\bibfnamefont {M.}~\bibnamefont
			{Dutra}}, \bibinfo {author} {\bibfnamefont {O.}~\bibnamefont
			{Louren\ifmmode~\mbox{\c{c}}\else \c{c}\fi{}o}}, \bibinfo {author}
		{\bibfnamefont {J.~S.}\ \bibnamefont {S\'a~Martins}}, \bibinfo {author}
		{\bibfnamefont {A.}~\bibnamefont {Delfino}}, \bibinfo {author} {\bibfnamefont
			{J.~R.}\ \bibnamefont {Stone}}, \ and\ \bibinfo {author} {\bibfnamefont
			{P.~D.}\ \bibnamefont {Stevenson}},\ }\href@noop {} {\bibfield  {journal}
		{\bibinfo  {journal} {Phys. Rev. C}\ }\textbf {\bibinfo {volume} {85}},\
		\bibinfo {pages} {035201} (\bibinfo {year} {2012})}\BibitemShut {NoStop}%
	\bibitem [{\citenamefont {Kortelainen}\ \emph {et~al.}(2010)\citenamefont
		{Kortelainen}, \citenamefont {Lesinski}, \citenamefont {Mor\'e},
		\citenamefont {Nazarewicz}, \citenamefont {Sarich}, \citenamefont {Schunck},
		\citenamefont {Stoitsov},\ and\ \citenamefont {Wild}}]{UNEDF0}%
	\BibitemOpen
	\bibfield  {author} {\bibinfo {author} {\bibfnamefont {M.}~\bibnamefont
			{Kortelainen}}, \bibinfo {author} {\bibfnamefont {T.}~\bibnamefont
			{Lesinski}}, \bibinfo {author} {\bibfnamefont {J.}~\bibnamefont {Mor\'e}},
		\bibinfo {author} {\bibfnamefont {W.}~\bibnamefont {Nazarewicz}}, \bibinfo
		{author} {\bibfnamefont {J.}~\bibnamefont {Sarich}}, \bibinfo {author}
		{\bibfnamefont {N.}~\bibnamefont {Schunck}}, \bibinfo {author} {\bibfnamefont
			{M.~V.}\ \bibnamefont {Stoitsov}}, \ and\ \bibinfo {author} {\bibfnamefont
			{S.}~\bibnamefont {Wild}},\ }\href@noop {} {\bibfield  {journal} {\bibinfo
			{journal} {Phys. Rev. C}\ }\textbf {\bibinfo {volume} {82}},\ \bibinfo
		{pages} {024313} (\bibinfo {year} {2010})}\BibitemShut {NoStop}%
	\bibitem [{\citenamefont {Kortelainen}\ \emph {et~al.}(2012)\citenamefont
		{Kortelainen}, \citenamefont {McDonnell}, \citenamefont {Nazarewicz},
		\citenamefont {Reinhard}, \citenamefont {Sarich}, \citenamefont {Schunck},
		\citenamefont {Stoitsov},\ and\ \citenamefont {Wild}}]{UNEDF1}%
	\BibitemOpen
	\bibfield  {author} {\bibinfo {author} {\bibfnamefont {M.}~\bibnamefont
			{Kortelainen}}, \bibinfo {author} {\bibfnamefont {J.}~\bibnamefont
			{McDonnell}}, \bibinfo {author} {\bibfnamefont {W.}~\bibnamefont
			{Nazarewicz}}, \bibinfo {author} {\bibfnamefont {P.-G.}\ \bibnamefont
			{Reinhard}}, \bibinfo {author} {\bibfnamefont {J.}~\bibnamefont {Sarich}},
		\bibinfo {author} {\bibfnamefont {N.}~\bibnamefont {Schunck}}, \bibinfo
		{author} {\bibfnamefont {M.~V.}\ \bibnamefont {Stoitsov}}, \ and\ \bibinfo
		{author} {\bibfnamefont {S.~M.}\ \bibnamefont {Wild}},\ }\href@noop {}
	{\bibfield  {journal} {\bibinfo  {journal} {Phys. Rev. C}\ }\textbf {\bibinfo
			{volume} {85}},\ \bibinfo {pages} {024304} (\bibinfo {year}
		{2012})}\BibitemShut {NoStop}%
	\bibitem [{\citenamefont {Kortelainen}\ \emph {et~al.}(2014)\citenamefont
		{Kortelainen}, \citenamefont {McDonnell}, \citenamefont {Nazarewicz},
		\citenamefont {Olsen}, \citenamefont {Reinhard}, \citenamefont {Sarich},
		\citenamefont {Schunck}, \citenamefont {Wild}, \citenamefont {Davesne},
		\citenamefont {Erler},\ and\ \citenamefont {Pastore}}]{UNEDF2}%
	\BibitemOpen
	\bibfield  {author} {\bibinfo {author} {\bibfnamefont {M.}~\bibnamefont
			{Kortelainen}}, \bibinfo {author} {\bibfnamefont {J.}~\bibnamefont
			{McDonnell}}, \bibinfo {author} {\bibfnamefont {W.}~\bibnamefont
			{Nazarewicz}}, \bibinfo {author} {\bibfnamefont {E.}~\bibnamefont {Olsen}},
		\bibinfo {author} {\bibfnamefont {P.-G.}\ \bibnamefont {Reinhard}}, \bibinfo
		{author} {\bibfnamefont {J.}~\bibnamefont {Sarich}}, \bibinfo {author}
		{\bibfnamefont {N.}~\bibnamefont {Schunck}}, \bibinfo {author} {\bibfnamefont
			{S.~M.}\ \bibnamefont {Wild}}, \bibinfo {author} {\bibfnamefont
			{D.}~\bibnamefont {Davesne}}, \bibinfo {author} {\bibfnamefont
			{J.}~\bibnamefont {Erler}}, \ and\ \bibinfo {author} {\bibfnamefont
			{A.}~\bibnamefont {Pastore}},\ }\href@noop {} {\bibfield  {journal} {\bibinfo
			{journal} {Phys. Rev. C}\ }\textbf {\bibinfo {volume} {89}},\ \bibinfo
		{pages} {054314} (\bibinfo {year} {2014})}\BibitemShut {NoStop}%
	\bibitem [{\citenamefont {Lim}\ and\ \citenamefont {Schwenk}(2024)}]{lim2024a}%
	\BibitemOpen
	\bibfield  {author} {\bibinfo {author} {\bibfnamefont {Y.}~\bibnamefont
			{Lim}}\ and\ \bibinfo {author} {\bibfnamefont {A.}~\bibnamefont {Schwenk}},\
	}\href@noop {} {\bibfield  {journal} {\bibinfo  {journal} {Phys. Rev. C}\
		}\textbf {\bibinfo {volume} {109}},\ \bibinfo {pages} {035801} (\bibinfo
		{year} {2024})}\BibitemShut {NoStop}%
	\bibitem [{\citenamefont {Hebeler}\ and\ \citenamefont
		{Schwenk}(2010)}]{Hebeler10}%
	\BibitemOpen
	\bibfield  {author} {\bibinfo {author} {\bibfnamefont {K.}~\bibnamefont
			{Hebeler}}\ and\ \bibinfo {author} {\bibfnamefont {A.}~\bibnamefont
			{Schwenk}},\ }\href@noop {} {\bibfield  {journal} {\bibinfo  {journal} {Phys.
				Rev. C}\ }\textbf {\bibinfo {volume} {82}},\ \bibinfo {pages} {014314}
		(\bibinfo {year} {2010})}\BibitemShut {NoStop}%
	\bibitem [{\citenamefont {Hagen}\ \emph
		{et~al.}(2014{\natexlab{b}})\citenamefont {Hagen}, \citenamefont
		{Papenbrock}, \citenamefont {Ekstr{\"o}m}, \citenamefont {Wendt},
		\citenamefont {Baardsen}, \citenamefont {Gandolfi}, \citenamefont
		{Hjorth-Jensen},\ and\ \citenamefont {Horowitz}}]{hagen14a}%
	\BibitemOpen
	\bibfield  {author} {\bibinfo {author} {\bibfnamefont {G.}~\bibnamefont
			{Hagen}}, \bibinfo {author} {\bibfnamefont {T.}~\bibnamefont {Papenbrock}},
		\bibinfo {author} {\bibfnamefont {A.}~\bibnamefont {Ekstr{\"o}m}}, \bibinfo
		{author} {\bibfnamefont {K.}~\bibnamefont {Wendt}}, \bibinfo {author}
		{\bibfnamefont {G.}~\bibnamefont {Baardsen}}, \bibinfo {author}
		{\bibfnamefont {S.}~\bibnamefont {Gandolfi}}, \bibinfo {author}
		{\bibfnamefont {M.}~\bibnamefont {Hjorth-Jensen}}, \ and\ \bibinfo {author}
		{\bibfnamefont {C.}~\bibnamefont {Horowitz}},\ }\href@noop {} {\bibfield
		{journal} {\bibinfo  {journal} {Phys. Rev. C}\ }\textbf {\bibinfo {volume}
			{89}},\ \bibinfo {pages} {014319} (\bibinfo {year}
		{2014}{\natexlab{b}})}\BibitemShut {NoStop}%
	\bibitem [{\citenamefont {Holt}\ and\ \citenamefont
		{Kaiser}(2017)}]{Holt2016pjb}%
	\BibitemOpen
	\bibfield  {author} {\bibinfo {author} {\bibfnamefont {J.~W.}\ \bibnamefont
			{Holt}}\ and\ \bibinfo {author} {\bibfnamefont {N.}~\bibnamefont {Kaiser}},\
	}\href@noop {} {\bibfield  {journal} {\bibinfo  {journal} {Phys. Rev.}\
		}\textbf {\bibinfo {volume} {C95}},\ \bibinfo {pages} {034326} (\bibinfo
		{year} {2017})}\BibitemShut {NoStop}%
	\bibitem [{\citenamefont {Stroberg}\ \emph {et~al.}(2021)\citenamefont
		{Stroberg}, \citenamefont {Holt}, \citenamefont {Schwenk},\ and\
		\citenamefont {Simonis}}]{stroberg21}%
	\BibitemOpen
	\bibfield  {author} {\bibinfo {author} {\bibfnamefont {S.~R.}\ \bibnamefont
			{Stroberg}}, \bibinfo {author} {\bibfnamefont {J.~D.}\ \bibnamefont {Holt}},
		\bibinfo {author} {\bibfnamefont {A.}~\bibnamefont {Schwenk}}, \ and\
		\bibinfo {author} {\bibfnamefont {J.}~\bibnamefont {Simonis}},\ }\href@noop
	{} {\bibfield  {journal} {\bibinfo  {journal} {Phys. Rev. Lett.}\ }\textbf
		{\bibinfo {volume} {126}},\ \bibinfo {pages} {022501} (\bibinfo {year}
		{2021})}\BibitemShut {NoStop}%
	\bibitem [{\citenamefont {Neufcourt}\ \emph {et~al.}(2019)\citenamefont
		{Neufcourt}, \citenamefont {Cao}, \citenamefont {Nazarewicz}, \citenamefont
		{Olsen},\ and\ \citenamefont {Viens}}]{neufcourt19}%
	\BibitemOpen
	\bibfield  {author} {\bibinfo {author} {\bibfnamefont {L.}~\bibnamefont
			{Neufcourt}}, \bibinfo {author} {\bibfnamefont {Y.}~\bibnamefont {Cao}},
		\bibinfo {author} {\bibfnamefont {W.}~\bibnamefont {Nazarewicz}}, \bibinfo
		{author} {\bibfnamefont {E.}~\bibnamefont {Olsen}}, \ and\ \bibinfo {author}
		{\bibfnamefont {F.}~\bibnamefont {Viens}},\ }\href@noop {} {\bibfield
		{journal} {\bibinfo  {journal} {Phys. Rev. Lett.}\ }\textbf {\bibinfo
			{volume} {122}},\ \bibinfo {pages} {062502} (\bibinfo {year}
		{2019})}\BibitemShut {NoStop}%
	\bibitem [{\citenamefont {Baym}\ \emph {et~al.}(1971)\citenamefont {Baym},
		\citenamefont {Bethe},\ and\ \citenamefont {Pethick}}]{baym71}%
	\BibitemOpen
	\bibfield  {author} {\bibinfo {author} {\bibfnamefont {G.}~\bibnamefont
			{Baym}}, \bibinfo {author} {\bibfnamefont {H.~A.}\ \bibnamefont {Bethe}}, \
		and\ \bibinfo {author} {\bibfnamefont {C.~J.}\ \bibnamefont {Pethick}},\
	}\href@noop {} {\bibfield  {journal} {\bibinfo  {journal} {Nucl. Phys. A}\
		}\textbf {\bibinfo {volume} {175}},\ \bibinfo {pages} {225} (\bibinfo {year}
		{1971})}\BibitemShut {NoStop}%
	\bibitem [{\citenamefont {Pethick}\ \emph {et~al.}(1995)\citenamefont
		{Pethick}, \citenamefont {Ravenhall},\ and\ \citenamefont
		{Lorenz}}]{pethick95}%
	\BibitemOpen
	\bibfield  {author} {\bibinfo {author} {\bibfnamefont {C.}~\bibnamefont
			{Pethick}}, \bibinfo {author} {\bibfnamefont {D.}~\bibnamefont {Ravenhall}},
		\ and\ \bibinfo {author} {\bibfnamefont {C.}~\bibnamefont {Lorenz}},\
	}\href@noop {} {\bibfield  {journal} {\bibinfo  {journal} {Nucl. Phys. A}\
		}\textbf {\bibinfo {volume} {584}},\ \bibinfo {pages} {675} (\bibinfo {year}
		{1995})}\BibitemShut {NoStop}%
	\bibitem [{\citenamefont {Kubis}(2007)}]{kubis07}%
	\BibitemOpen
	\bibfield  {author} {\bibinfo {author} {\bibfnamefont {S.}~\bibnamefont
			{Kubis}},\ }\href@noop {} {\bibfield  {journal} {\bibinfo  {journal} {Phys.
				Rev. C}\ }\textbf {\bibinfo {volume} {76}},\ \bibinfo {pages} {025801}
		(\bibinfo {year} {2007})}\BibitemShut {NoStop}%
	\bibitem [{\citenamefont {Lattimer}\ and\ \citenamefont
		{Prakash}(2007)}]{lattimer07rep}%
	\BibitemOpen
	\bibfield  {author} {\bibinfo {author} {\bibfnamefont {J.~M.}\ \bibnamefont
			{Lattimer}}\ and\ \bibinfo {author} {\bibfnamefont {M.}~\bibnamefont
			{Prakash}},\ }\href@noop {} {\bibfield  {journal} {\bibinfo  {journal}
			{Physics Reports}\ }\textbf {\bibinfo {volume} {442}},\ \bibinfo {pages}
		{109} (\bibinfo {year} {2007})}\BibitemShut {NoStop}%
	\bibitem [{\citenamefont {Hebeler}\ \emph {et~al.}(2013)\citenamefont
		{Hebeler}, \citenamefont {Lattimer}, \citenamefont {Pethick},\ and\
		\citenamefont {Schwenk}}]{hebeler13}%
	\BibitemOpen
	\bibfield  {author} {\bibinfo {author} {\bibfnamefont {K.}~\bibnamefont
			{Hebeler}}, \bibinfo {author} {\bibfnamefont {J.~M.}\ \bibnamefont
			{Lattimer}}, \bibinfo {author} {\bibfnamefont {C.~J.}\ \bibnamefont
			{Pethick}}, \ and\ \bibinfo {author} {\bibfnamefont {A.}~\bibnamefont
			{Schwenk}},\ }\href@noop {} {\bibfield  {journal} {\bibinfo  {journal}
			{Astrophys. J.}\ }\textbf {\bibinfo {volume} {773}},\ \bibinfo {pages} {11}
		(\bibinfo {year} {2013})}\BibitemShut {NoStop}%
	\bibitem [{\citenamefont {Holt}\ \emph
		{et~al.}(2013{\natexlab{b}})\citenamefont {Holt}, \citenamefont {Kaiser},\
		and\ \citenamefont {Weise}}]{PhysRevC.87.014338}%
	\BibitemOpen
	\bibfield  {author} {\bibinfo {author} {\bibfnamefont {J.~W.}\ \bibnamefont
			{Holt}}, \bibinfo {author} {\bibfnamefont {N.}~\bibnamefont {Kaiser}}, \ and\
		\bibinfo {author} {\bibfnamefont {W.}~\bibnamefont {Weise}},\ }\href@noop {}
	{\bibfield  {journal} {\bibinfo  {journal} {Phys. Rev. C}\ }\textbf {\bibinfo
			{volume} {87}},\ \bibinfo {pages} {014338} (\bibinfo {year}
		{2013}{\natexlab{b}})}\BibitemShut {NoStop}%
	\bibitem [{\citenamefont {Chabanat}\ \emph {et~al.}(1997)\citenamefont
		{Chabanat}, \citenamefont {Bonche}, \citenamefont {Haensel}, \citenamefont
		{Meyer},\ and\ \citenamefont {Schaeffer}}]{Chabanat1997}%
	\BibitemOpen
	\bibfield  {author} {\bibinfo {author} {\bibfnamefont {E.}~\bibnamefont
			{Chabanat}}, \bibinfo {author} {\bibfnamefont {P.}~\bibnamefont {Bonche}},
		\bibinfo {author} {\bibfnamefont {P.}~\bibnamefont {Haensel}}, \bibinfo
		{author} {\bibfnamefont {J.}~\bibnamefont {Meyer}}, \ and\ \bibinfo {author}
		{\bibfnamefont {R.}~\bibnamefont {Schaeffer}},\ }\href@noop {} {\bibfield
		{journal} {\bibinfo  {journal} {Nuclear Physics A}\ }\textbf {\bibinfo
			{volume} {627}},\ \bibinfo {pages} {710} (\bibinfo {year}
		{1997})}\BibitemShut {NoStop}%
	\bibitem [{\citenamefont {Rikovska~Stone}\ \emph {et~al.}(2003)\citenamefont
		{Rikovska~Stone}, \citenamefont {Miller}, \citenamefont {Koncewicz},
		\citenamefont {Stevenson},\ and\ \citenamefont {Strayer}}]{Stone2003}%
	\BibitemOpen
	\bibfield  {author} {\bibinfo {author} {\bibfnamefont {J.}~\bibnamefont
			{Rikovska~Stone}}, \bibinfo {author} {\bibfnamefont {J.~C.}\ \bibnamefont
			{Miller}}, \bibinfo {author} {\bibfnamefont {R.}~\bibnamefont {Koncewicz}},
		\bibinfo {author} {\bibfnamefont {P.~D.}\ \bibnamefont {Stevenson}}, \ and\
		\bibinfo {author} {\bibfnamefont {M.~R.}\ \bibnamefont {Strayer}},\
	}\href@noop {} {\bibfield  {journal} {\bibinfo  {journal} {Phys. Rev. C}\
		}\textbf {\bibinfo {volume} {68}},\ \bibinfo {pages} {034324} (\bibinfo
		{year} {2003})}\BibitemShut {NoStop}%
	\bibitem [{\citenamefont {Goriely}\ \emph {et~al.}(2005)\citenamefont
		{Goriely}, \citenamefont {Samyn}, \citenamefont {Pearson},\ and\
		\citenamefont {Onsi}}]{Goriely2005}%
	\BibitemOpen
	\bibfield  {author} {\bibinfo {author} {\bibfnamefont {S.}~\bibnamefont
			{Goriely}}, \bibinfo {author} {\bibfnamefont {M.}~\bibnamefont {Samyn}},
		\bibinfo {author} {\bibfnamefont {J.}~\bibnamefont {Pearson}}, \ and\
		\bibinfo {author} {\bibfnamefont {M.}~\bibnamefont {Onsi}},\ }\href@noop {}
	{\bibfield  {journal} {\bibinfo  {journal} {Nuclear Physics A}\ }\textbf
		{\bibinfo {volume} {750}},\ \bibinfo {pages} {425} (\bibinfo {year}
		{2005})}\BibitemShut {NoStop}%
\end{thebibliography}
%

\end{document}